\documentclass[floats,aps,amssymb,nofootinbib,showkeys]{revtex4}
\usepackage{amssymb}
\usepackage{graphics}
\usepackage{amsmath}
\usepackage{slashed}
\usepackage{xcolor}
\usepackage{amsfonts}
\usepackage{bm}
\usepackage[]{latexsym}
\DeclareMathOperator{\sech}{sech}
\DeclareMathOperator{\csch}{csch}

\newcommand{\be}{\begin{equation}}\newcommand{\ee}{\end{equation}}
\newcommand{\bea}{\begin{eqnarray}}\newcommand{\eea}{\end{eqnarray}}
\newcommand{\brr}{\begin{array}}\newcommand{\err}{\end{array}}
\newcommand{\bit}{\begin{itemize}}\newcommand{\eit}{\end{itemize}}
\newcommand{\ben}{\begin{enumerate}}\newcommand{\een}{\end{enumerate}}

\newcommand{\ba}{\begin{array}}
\newcommand{\ea}{\end{array}}

\def\lf{\left}

\def\non{\nonumber}

\def\ri{\right}

\def\1{{_{1}}}\def\2{{_{2}}}
\def\bk{{\bf {k}}}\def\bx{{\bf {x}}}

\def\noHe0{:\;\!\!\;\!\!:H_e(0):\;\!\!\;\!\!:}
\def\noHm0{:\;\!\!\;\!\!:H_\mu(0):\;\!\!\;\!\!:}

\def\lf{\left}

\def\non{\nonumber}

\def\ri{\right}

\def\1{{_{1}}}\def\2{{_{2}}}

\def\lf{\left}

\def\non{\nonumber}

\def\ri{\right}

\def\1{{_{1}}}\def\2{{_{2}}}

\begin{document}
\title{Nonextensive Tsallis statistics in Unruh effect for Dirac neutrinos}

\author{Giuseppe Gaetano Luciano\footnote{email: gluciano@sa.infn.it (corresponding author)}$^{\hspace{0.1mm}1,2}$ and Massimo Blasone\footnote{email: blasone@sa.infn.it}$^{\hspace{0.1mm}1,2}$} 
\affiliation
{$^1$Dipartimento di Fisica, Universit\`a di Salerno, Via Giovanni Paolo II, 132 I-84084 Fisciano (SA), Italy.\\ 
$^2$INFN, Sezione di Napoli, Gruppo collegato di Salerno, Via Giovanni Paolo II, 132 I-84084 Fisciano (SA), Italy.}

\date{\today}

\begin{abstract}
Flavor mixing of quantum fields was found to be responsible
for the breakdown of the thermal Unruh effect. 
Recently, this result was revisited in the context of 
nonextensive Tsallis thermostatistics, showing that the emergent vacuum condensate
can still be featured as a thermal-like bath,
provided that the underlying statistics is assumed to obey Tsallis prescription. This was analyzed explicitly for bosons. Here we extend this study to Dirac fermions and in particular to neutrinos. Working in the relativistic approximation, we provide an effective description of the 
modified Unruh spectrum in terms of the $q$-generalized Tsallis statistics, 
the $q$-entropic index being dependent
on the mixing parameters $\sin\theta$ and $\Delta m$. 
As opposed to bosons, we find $q>1$, which is indicative of the subadditivity regime of Tsallis entropy. An intuitive understanding of this result is discussed
in relation to the nontrivial entangled structure exhibited by the quantum vacuum for mixed fields, combined with the Pauli exclusion principle.

\end{abstract}


\date{\today}
\setcounter{equation}{0}
\maketitle

\section{Introduction}
The Unruh effect states that an observer
moving through the inertial vacuum with uniform  
acceleration $a$ (Rindler observer) perceives a thermal radiation at temperature~\cite{Unruh}.
\be
\label{TU}
T_{\mathrm{U}}=a/2\pi\,,
\ee
in natural units.
Since its theoretical prediction, such an effect has stimulated deep
investigation, including applications to quantum optics
and quantum information~\cite{QO,Fuentes,Alsing,Adesso}, connections with cosmological scenarios~\cite{Davies, Das, Gine} and the
interplay with other genuinely field theoretical phenomena~\cite{Interp1,Interp2,Interp4,Interp5,Interp6,Interp7,Interp8,Interp9,Interp10,Hammad} (see~\cite{Crispino} for a review). Furthermore, 
experimental proposals have been
suggested in a variety of contexts, ranging from
direct attempts in particle physics~\cite{Ng,Schu}
to analog models in water~\cite{Water,Water2} and condensed matter systems~\cite{Analog,Analog2,Analog3}.
In spite of these efforts, the problem of revealing the Unruh radiation is still open, 
and several questions on its very existence have been raised during the years~\cite{Fedotov}.

Recently, the growing interest in the Unruh effect has led to unveil  unconventional features of this phenomenon. For instance, in~\cite{Marino} it was argued that the Casimir-Polder forces between two relativistic uniformly accelerated atoms exhibit a transition from the short distance thermal-like behavior predicted by the Unruh effect to a long distance non-thermal behavior. Similarly, in~\cite{Hammad} the Unruh temperature $T_{\mathrm{U}}$ was claimed to be sensitive to the degree of departure from linearity of the dispersion relation
for the case of massive fields. 
Exotic behaviors of Unruh radiation also appear when considering the superposition of fields with different masses, as stated in~\cite{Luciano,NonTN}. Specifically, in~\cite{Luciano} it was shown that the vacuum density of mixed particles detected by the Rindler observer deviates from the pure Planckian spectrum, the correction being quantified by the mass difference and the mixing angle. 
In turn, this feature can be ascribed to the peculiar nature of the vacuum for mixed fields (flavor vacuum), which appears as a condensate of entangled particle/antiparticle pairs due to the nontrivial structure of the mixing transformations at level of the ladder operators~\cite{BV95,BlasCap}. 

The mixing-induced departure of Unruh condensate from the Planckian
profile was originally interpreted as a breakdown of the thermality 
of Unruh effect~\cite{Luciano}. In a recent work~\cite{qLuciano}, this result
has been revisited in the context of the $q$-generalized Tsallis statistics based on the nonadditive Tsallis entropy~\cite{Tsallis1,Tsallis2,Tsallis3,Tsallis4}. As well-known, Tsallis entropy is a generalization of the standard Boltzmann-Gibbs definition parameterized by the $q$-index (in particular, $q\neq1$ corresponds to the nonextensive Tsallis statistics, while $q\rightarrow1$ gives back Boltzmann-Gibbs framework). The concept was introduced as a basis for extending the usual statistical mechanics and is formally identical to Havrda-Charv\'at structural $\alpha$-entropy considered in the information theory~\cite{Infth}. The rationale behind the study of the modified Unruh effect within Tsallis statistics is that the $q$-generalized entropy well describes systems exhibiting long-range interactions and/or spacetime entanglement, as is the case for mixed fields~\cite{BV95,BlasCap,Cabo,Vacent}. In this vein, 
in~\cite{qLuciano} it has been shown that the Unruh condensate for mixed particles can still be described as a thermal-like distribution, 
provided that the underlying statistics is assumed to obey Tsallis's prescription. This picture allows to relate the $q$-entropic index and the characteristic mixing parameters $\sin\theta$ and $\Delta m$ in a nontrivial way. Clearly, in the absence of mixing (i.e. for vanishing $\theta$ and/or $\Delta m$), the Boltzmann-Gibbs theory is recovered. In passing, we mention that 
applications of Tsallis entropy have been widely considered in 
literature. Among the various systems that clarify the physical conditions under which Tsallis entropy and the associated statistics apply, we quote self-gravitating stellar systems~\cite{App1,App3}, black holes~\cite{Tsallis3}, the cosmic background radiation~\cite{App7,App8}, low-dimensional dissipative systems~\cite{Tsallis4}, solar neutrinos~\cite{App11}, polymer chains~\cite{Polch} and cosmological models~\cite{App13,App14}. 

The description of Unruh effect in Tsallis's theory has been developed for mixing of boson fields. In that case, it has been found that $q<1$, 
indicating a superadditive behavior of
Tsallis entropy. In what follows, we extend this study
to the case of Dirac fermions and in particular to neutrinos, 
which are the paradigmatic example of mixed particles.
Again, we relate the $q$-index to the mixing angle
and the mass difference, in such a way that
the standard Fermi-Dirac distribution based on the extensive
Boltzmann-Gibbs entropy is restored for vanishing mixing. 
Similarly to~\cite{qLuciano}, we find a running behavior
of $q$ as a function of the energy scale, which is typical for QFT systems~\cite{App13}.
Nevertheless, as opposed to bosons, we now obtain 
$q>1$, that is to say, the entropy function turns out to be
subadditive. Following~\cite{qLuciano}, we explore
our result in connection with the entangled condensate structure
exhibited by the quantum vacuum for mixed fields. 

The remainder of the work is organized as follows: in Sec.~\ref{UE}
we review the quantization of Dirac field in Rindler spacetime and
the origin of the Unruh effect. Section~\ref{QFTRIND} is devoted to the
study of the QFT formalism of field mixing in Rindler spacetime.
This prepares the ground for the analysis of Sec.~\ref{EDM}, 
where we introduce the framework of Tsallis thermodynamics 
and investigate the connection with the phenomenon of field mixing.
To avoid technicalities, we make use of some approximations, such as those of relativistic neutrinos and 
small deviations of $q$ from unity. 
Discussion of the results and outlook are summarized in Sec.~\ref{Conc}.
Two Appendices containing details of calculations 
conclude the work.
Throughout all the manuscript, we use the $4$-dimensional metric 
with the mostly negative signature 
\be
\eta_{\mu\nu}=(+1,-1,-1,-1)\,.
\ee
Furthermore we adopt natural units $\hbar=c=k_{\mathrm{B}}=1$.

\maketitle

\section{Field quantization in Rindler spacetime: the Unruh effect}
\label{UE}
Let us consider an observer moving in Minkowski spacetime
along the $x$-axis with constant proper acceleration $a>0$. From~\cite{Mukhanov},
it is known that his trajectory is given by
\be
\label{traj}
t=a^{-1}\sinh(a\tau), \qquad x=a^{-1}\cosh(a\tau), \qquad y=y(0), \qquad z=z(0), 
\ee
parameterized by $\tau$. It is easy to see that this is 
an hyperbola in the $(t,x)$-plane of asymptotes $x=\pm t$, 
which act as event horizons for the accelerated observer. 
By varying $a$, we obtain different hyperbolas with
the same geometric (and, thus, physical) properties. 

Starting from the Minkowski metric $ds^2=dt^2-dx^2-dy^2-dz^2$, 
we now introduce the following set of coordinates
\begin{eqnarray}
\label{coord}
&t=\rho\sinh\eta,\qquad x=\rho\cosh\eta,&\\[2mm]
&\rho=\sqrt{x^2-t^2},\qquad \eta=\mathrm{arctgh}\left(t/x\right),&
\label{coord2}
\end{eqnarray}
with the other coordinates left unchanged. The metric then becomes
\be
\label{rind}
ds^2=\rho^2d\eta^2-d\rho^2-dy^2-dz^2\,,
\ee
which describes a stationary spacetime of
timelike Killing vector $\partial/\partial \eta$. 

\begin{figure}[t]
\resizebox{8.5cm}{!}{\includegraphics{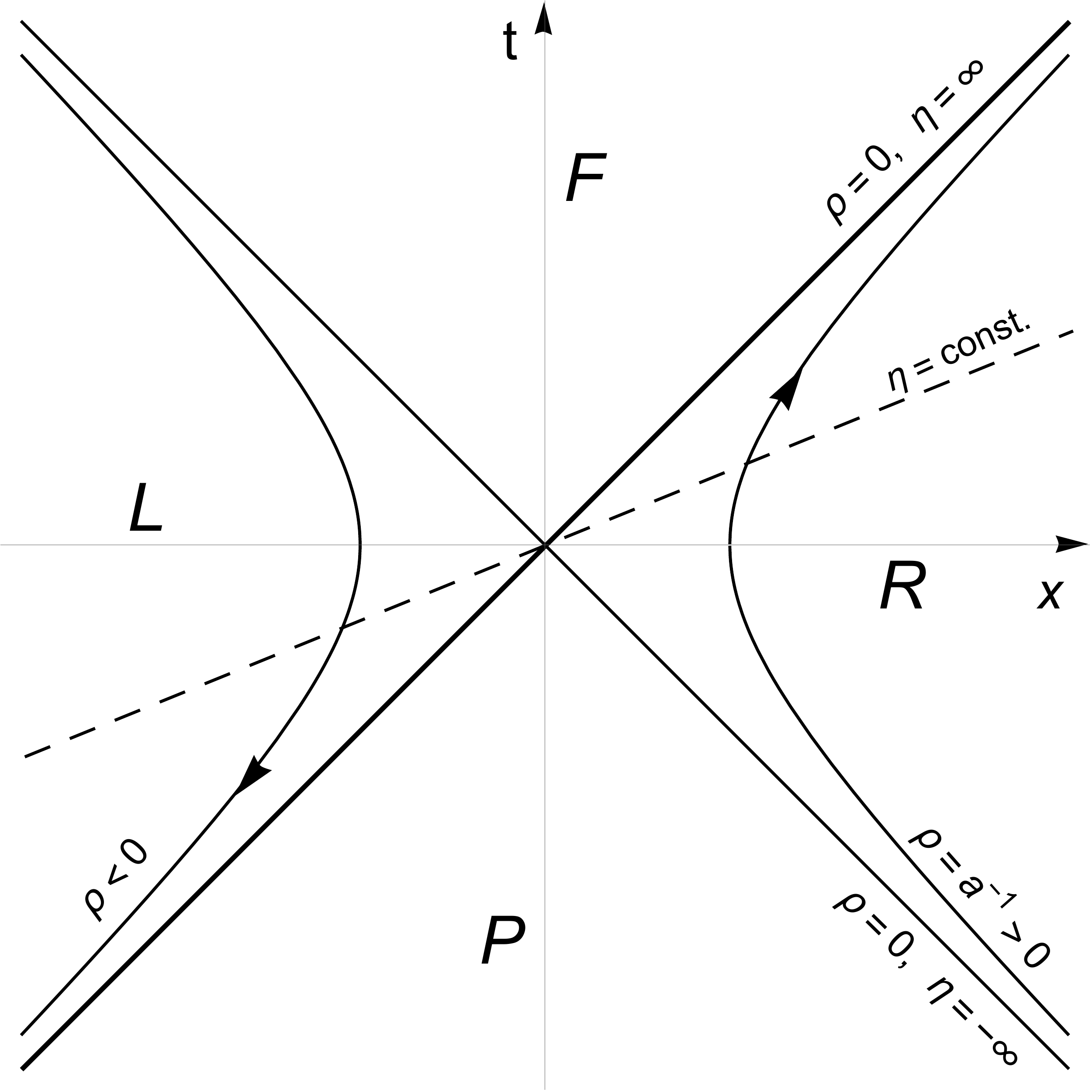}}
\caption{\small{Causal structure of Rindler metric in $1+1$ dimensions. 
We are assuming that the acceleration is along the $x$-axis. The right branch of hyperbola represents the worldine of the Rindler observer. The arrow indicates the direction of motion.}}
\label{rindfig}
\end{figure}

By comparing Eqs.~\eqref{traj} and~\eqref{coord}, 
the wordlines $\rho=\mathrm{const}$, $y=\mathrm{const}$, 
$z=\mathrm{const}$ correspond to the trajectories
of uniformly accelerated observers with $a=\rho^{-1}$ and
proper time $\tau=a\eta$. The collection of these worldlines 
forms the so-called Rindler spacetime (see the right $R$-wedge in 
Fig.~\ref{rindfig}). Notice that the
hypersurfaces of $\eta=\mathrm{const}$ describe
events which are simultaneous from the point of view
of the Rindler (uniformly accelerated) observer. 

In spite of the minimalistic setting, the geometric structure
of Rindler manifold differs significantly from that
of Minkowski spacetime. Indeed, an observer moving along a fixed hyperbola in $R$ is causally disconnected from the $L$-sector, which
means that he can neither send nor receive any signal from a source
in that region. Of course, this does not apply
to inertial (Minkowski) observers, which are indeed
connected to the entire spacetime. 

Strictly speaking, the set of coordinates~\eqref{coord}-\eqref{coord2} covers only the right sector of Minkowski spacetime. To describe the left $L$-wedge, one has
to require $\rho=-\sqrt{x^2-t^2}<0$, with $\eta$ running in the opposite direction. On the other hand, the future $F$ and past $P$ regions are covered by the charts
\begin{eqnarray}
&t=\rho\cosh\eta,\qquad x=\rho\sinh\eta,&\\[2mm]
&\rho=\pm\sqrt{t^2-x^2},\qquad \eta=\mathrm{arctgh}\left(x/t\right),&
\end{eqnarray}
where the $\pm$ signs refer to $F$ and $P$, respectively. We emphasize that these charts are obtained by inverting the r\^oles of time and space coordinates in Eq.~\eqref{coord}. For our next purposes, it is enough to consider
the right and left wedges only. 

\subsection{Minkowski spacetime}

Let us now turn to the problem of quantizing the Dirac
field in Rindler spacetime. Toward this end, it is useful to 
review the canonical planewave expansion in Minkowski space.
This is given by
\begin{equation}
\label{eqn:planewaveexpansion}
\Psi(t,\textbf{x}) =\sum_{r=1,2} \int d^{\hspace{0.2mm}3}\hspace{0.1mm}k\left[\hspace{0.2mm}a^r_{\hspace{0.2mm}\bk}\hspace{0.8mm}\psi^{r\hspace{0.4mm}+}_{\hspace{0.2mm\bk}}(t,\textbf{x})\,+\,b^{r\hspace{0.2mm}\dagger}_{\hspace{0.2mm}\bk}\hspace{0.8mm}\psi^{r\hspace{0.4mm}-}_{\hspace{0.2mm\bk}}(t,\textbf{x})\hspace{0.2mm}\right],
\end{equation}
where $\textbf{x}\equiv(x,y,z)$ denotes the set of spatial Minkowski coordinates. Here, $\psi^{r\hspace{0.4mm}+}_{\hspace{0.2mm\bk}}=N u^r_{\hspace{0.2mm}\bk}\hspace{0.4mm}e^{-i\omega_{\textbf{k}}t+i\textbf{k}\cdot\textbf{x}}$ and $\psi^{r\hspace{0.4mm}-}_{\hspace{0.2mm\bk}}=Nv^r_{\hspace{0.2mm}\bk}\hspace{0.4mm}e^{i\omega_{\textbf{k}}t-i\textbf{k}\cdot\textbf{x}}$ are the positive and negative frequency modes with three-momentum $\textbf{k}\equiv(k_x,k_y,k_z)$ and frequency $\omega_{\textbf{k}}=\sqrt{m^2+|\bk|^2}$, respectively. The normalization coefficient is fixed to $N=(2\pi)^{-3/2}$. 
The following explicit form of the spinors $u^r_{\textbf{k}}$ and $v^r_{\textbf{k}}$ is adopted 
\begin{equation}
u^r_\textbf{k}=\frac{\slashed p+m}{{\sqrt{2\omega_{\textbf{k}}\left(\omega_{\textbf{k}}+m\right)}}}\,u^r(0)\,,
\qquad v^r_{\textbf{k}}=\frac{-\slashed p+m}{{\sqrt{2\omega_{\textbf{k}}\left(\omega_{\textbf{k}}+m\right)}}}\,v^r(0)\,, 
\end{equation}
where we have used the usual notation $\slashed p=p^{\mu}\gamma_\mu$, with $\gamma^{\mu}$ being the gamma matrices in the Dirac representation and $u^r(0)$, $v^r(0)$ the unit spinors in the particle rest frame of $\textbf{k}=(m,0)$ (see~\cite{Greiner}). They are normalized according to
\be
u^{r\dagger}_\textbf{k}u^s_{\textbf{k}}=\delta_{rs}\,,
\ee
and similarly for $v^r_{\textbf{k}}$.

In the second quantization formalism
the operators $a^r_{\hspace{0.2mm}\bk}$ and $b^r_{\hspace{0.2mm}\bk}$ 
act as annihilators of the ordinary Minkowski vacuum $|0\rangle_{\mathrm{M}}$, i.e.
\be
\label{Minkvac}
a^{\hspace{0.2mm}r}_{\hspace{0.2mm}\bk}|0\rangle_{\mathrm{M}}=b^{\hspace{0.2mm}r}_{\hspace{0.2mm}\bk}|0\rangle_{\mathrm{M}}=0\,, \quad \forall \textbf{k}, r\,, 
\ee 
while $a^{r\dagger}_{\hspace{0.2mm}\bk}$ and $b^{r\dagger}_{\hspace{0.2mm}\bk}$ create a particle/antiparticle with quantum 
numbers $\textbf{k}$ and $r$. Such operators are assumed to satisfy the canonical anticommutation relations $ \left\{a^r_{\hspace{0.2mm}\bk}, a^{s\hspace{0.2mm}\dagger}_{\hspace{0.2mm}\bk'}\right\}=\left\{b^r_{\hspace{0.2mm}\bk}, b^{s\hspace{0.2mm}\dagger}_{\hspace{0.2mm}\bk'}\right\}=\delta_{rs}\,\delta\left(\textbf{k}-\textbf{k}'\right)$, with all other anticommutators vanishing. In turn, this implies the following relation for the equal-time anticommutator between the field and its hermitian conjugate
\be
\label{anticomcan}
\left\{\Psi_a(t,\textbf{x}),\Psi_\beta^\dagger(t,\textbf{x}')\right\}=\delta_{\alpha\beta}\,\delta^3\left(\textbf{x}-\textbf{x}'\right), 
\ee
where $\alpha$ and $\beta$ are the spinorial indices.

\subsection{Rindler spacetime}
We now properly aim at performing the second quantization
of the spinor field in Rindler spacetime. To fix the notation, we closely follow~\cite{Oriti} and consider
the solutions of the Dirac equation
in Rindler coordinates in the $R$-wedge. We emphasize that these
modes can be extended to the remaining 
sectors $F$, $L$ and $P$ via analytic continuation
across the event horizons, which represent branch points
for these functions. 

By using Eq.~\eqref{coord}, the Dirac
equation in $R$ takes the form
\be
\left(i\partial_\eta+i\rho\gamma^0\gamma^1\partial_\rho+
i\rho\gamma^0\gamma^2\partial_y+
i\rho\gamma^0\gamma^3\partial_z+
\frac{i}{2}\gamma^0\gamma^1-m\rho\gamma^0
\right)\psi(\eta,\rho,\vec{x})=0\,. 
\ee
This admits as solutions
\begin{equation}
\label{eqn:psirindcoord}
\Psi_{1,\kappa}(\eta,\rho,\vec{x})=N_\kappa\left(X_{1,\vec{k}}\,K_{i\Omega\,-\,\frac{1}{2}}\left(\mu_{\vec{k}}\hspace{0.2mm}\rho\right)\,+\,Y_{1,\vec{k}}\,K_{i\Omega\,+\,\frac{1}{2}}\left(\mu_{\vec{k}}\hspace{0.2mm}\rho\right)\right)e^{-i\hspace{0.2mm}\Omega\hspace{0.2mm}\eta}\hspace{0.3mm}e^{i\hspace{0.2mm}\vec{k}\cdot\hspace{0.2mm}\vec{x}}\,,
\end{equation}
with 
\be
X_{1,\vec{k}}=\begin{pmatrix}
  k_z \\[1mm]
  i(k_y+i m) \\[1mm]
  i(k_y+i m)\\[1mm]
  k_z \\
\end{pmatrix},\qquad
Y_{1,\vec{k}}=\begin{pmatrix}
  0 \\[1mm]
  i\mu_{\vec{k}} \\[1mm]
-i\mu_{\vec{k}}\\[1mm]
 0 \\
\end{pmatrix},
\ee
and
\begin{equation}
\label{eqn:psirindcoord2}
\Psi_{2,\kappa}(\eta,\rho,\vec{x})=N_\kappa\left(X_{2,\vec{k}}\,K_{i\Omega\,-\,\frac{1}{2}}\left(\mu_{\vec{k}}\hspace{0.2mm}\rho\right)\,+\,Y_{2,\vec{k}}\,K_{i\Omega\,+\,\frac{1}{2}}\left(\mu_{\vec{k}}\hspace{0.2mm}\rho\right)\right)e^{-i\hspace{0.2mm}\Omega\hspace{0.2mm}\eta}\hspace{0.3mm}e^{i\hspace{0.2mm}\vec{k}\cdot\hspace{0.2mm}\vec{x}}\,,
\end{equation}
with
\be
X_{2,\vec{k}}=\begin{pmatrix}
  0 \\[1mm]
  i\mu_{\vec{k}} \\[1mm]
  i\mu_{\vec{k}}\\[1mm]
  0\\
\end{pmatrix},\qquad
Y_{2,\vec{k}}=\begin{pmatrix}
  k_z \\[1mm]
  i(k_y-i m) \\[1mm]
 - i(k_y-i m)\\[1mm]
  -k_z \\
\end{pmatrix}.
\ee
In the above equations we have used the shorthand
notation $\vec{x}\equiv(y,z)$ and $\vec{k}\equiv (k_y,k_z)$ for the
transverse positon and momentum coordinates, respectively. 
The Hamiltonian eigenvalue
has been denoted by $\Omega>0$, while $\kappa\equiv\{\Omega,\vec{k}\}$. The $\rho$-dependence of the Rindler modes is contained in the Bessel function $K_{i\Omega\,\pm\,\frac{1}{2}}$, where $\mu_{\vec{k}}=\sqrt{m^2+|\vec{k}|^2}$ is the reduced frequency. 
The coefficient $N_\kappa={(4\hspace{0.2mm}\pi^2\hspace{0.2mm}{\sqrt{\mu_{\vec{k}}})}}^{-1}\,\sqrt{{\cosh{\left(\pi\hspace{0.2mm}\Omega\right)}}}$ provides the normalization respect to the inner product in Rindler coordinates~\cite{Oriti}.

For later purposes, it is also useful to consider
the solutions~\eqref{eqn:psirindcoord} and~\eqref{eqn:psirindcoord2}
expressed in Minkowski coordinates. By taking into
account the spinor transformations under coordinate changes, 
these solutions take the form~\cite{Oriti}
\be
\label{RindinMink}
\Psi_{j,\kappa}(t,\textbf{x})=N_\kappa
\left(X_{j,\vec{k}}\,K_{i\Omega\,-\,\frac{1}{2}}\left(\mu_{\vec{k}}\hspace{0.2mm}\rho\right)\,e^{-\left(i\hspace{0.2mm}\Omega-\frac{1}{2}\hspace{0.2mm}\right)\eta}
+Y_{j,\vec{k}}\,K_{i\Omega\,+\,\frac{1}{2}}\left(\mu_{\vec{k}}\hspace{0.2mm}\rho\right)\,e^{-\left(i\hspace{0.2mm}\Omega+\frac{1}{2}\hspace{0.2mm}\right)\eta}\hspace{0.3mm}\right) e^{i\hspace{0.2mm}\vec{k}\cdot\hspace{0.2mm}\vec{x}}\,,\qquad j=1,2\,.
\ee
We stress that these functions are still defined
in the $R$-wedge only.

Exploiting the above tools, we now expand the Dirac field in $R$ as
\be
\label{Rindfi}
\Psi_R(\eta,\rho,\vec{x})=\sum_{j=1,2}\int d^3\kappa
\left[a_{j,\kappa}\psi_{j,\kappa}(\eta,\rho,\vec{x}) +b^\dagger_{j,\kappa}\psi_{j,\tilde\kappa}(\eta,\rho,\vec{x})
\right],
\ee
where $\tilde\kappa\equiv(-\Omega,\vec{k})$. As
for the Minkowski quantization, the operators
$a_{j,\kappa}$ and $b_{j,\kappa}$
are assumed to be canonical, i.e.
\be
\left\{a_{j,\kappa},a^\dagger_{i,\kappa'}\right\}=
\left\{b_{j,\kappa},b^\dagger_{i\kappa'}\right\}=
\delta_{ji}\delta^3\left(\kappa-\kappa'\right)\,,
\ee
with all other anticommutators being trivial. 
They now annihilate the Rindler vacuum state
\be
a_{j,\kappa}|0\rangle_{\mathrm{R}}=b_{j,\kappa}|0\rangle_{\mathrm{R}}=0\,, \quad \forall \kappa, j\,, 
\ee 
while $a^\dagger_{j,\kappa}$ and $b^\dagger_{j,\kappa}$
create a particle/antiparticle for the Rindler observer in $R$. 

To make it easier the comparison between the 
field expansions~\eqref{eqn:planewaveexpansion} and~\eqref{Rindfi}, 
let us introduce an alternative field quantization
in Minkowski spacetime, namely a quantization in terms
of the Lorentz momentum eigenfunctions. The physical
reason behind this formalism is that the Rindler Hamiltonian
is exactly the Lorentz boost generator written in Minkowski coordinates.
This means that the time evolution of a Rindler observer
can be properly described as an infinite succession
of infinitesimal boost transformations (in our case along the $x$-axis).

We consider the integral representations of the 
Bessel function $K_\nu(\rho)$ given by
\begin{eqnarray}
\label{firep}
K_{\nu}(\rho)&=&\frac{1}{2}e^{-\frac{i\pi\nu}{2}}\int_{-\infty}^{\infty}d\theta\, e^{i\rho\sinh\theta}e^{\nu\theta}\,,\\[2mm]
K_{\nu}(\rho)&=&\frac{1}{2}e^{\frac{i\pi\nu}{2}}\int_{-\infty}^{\infty}d\theta\, e^{i\rho\sinh\theta}e^{-\nu\theta}\,.
\end{eqnarray}
From the last representation, by using the coordinate
transformation~\eqref{coord} one can show that~\cite{Oriti}
\be
\label{posfr}
K_{\nu}(\mu_{\vec{k}}\hspace{0.5mm}\rho(t,x))\,e^{-\nu\eta}
=\frac{1}{2}e^{\frac{i\pi\nu}{2}}\int_{-\infty}^{\infty}d\theta\, P_\theta^-(t,x)\, e^{-\nu\theta}\,,
\ee
where $P_\theta^-(t,x)$ is the two-dimensional 
positive-frequency planewave with  
$\omega=\mu_{\vec{k}}\cosh\theta$
and $k_x=\mu_{\vec{k}}\sinh\theta$. On the other hand, 
Eq.~\eqref{firep} leads to
\be
\label{negfr}
K_{\nu}(\mu_{\vec{k}}\hspace{0.5mm}\rho(t,x))\,e^{-\nu\eta}
=\frac{1}{2}e^{-\frac{i\pi\nu}{2}}\int_{-\infty}^{\infty}d\theta\, P_\theta^+(t,x)\, e^{\nu\theta}\,,
\ee
where $P_\theta^+(t,x)$ is the two-dimensional 
negative-frequency planewave. Remarkably, 
Eqs.~\eqref{posfr} and~\eqref{negfr}
show that the function $K_{\nu}(\mu_{\vec{k}}\hspace{0.5mm}\rho(t,x))\,e^{-\nu\eta}$ can be expressed equivalently
as superposition of either positive- or negative-frequency
planewaves. 

If we now plug the above representations of the Bessel function into 
Eq.~\eqref{RindinMink}, we obtain
\begin{equation}
\label{global}
\psi^{\mp}_{j,\kappa}(t,\textbf{x})=\frac{1}{2}N^{\mp}_\kappa
\left[X_{j,\vec{k}}\,e^{\pm \frac{i\pi}{2}\left(i\Omega-\frac{1}{2}\right)}
\int_{-\infty}^{\infty}d\theta\, P^{\mp}_\theta(t,x)\,e^{\mp\left(i\Omega-\frac{1}{2}\right)\theta}
+
Y_{j,\vec{k}}\,e^{\pm \frac{i\pi}{2}\left(i\Omega+\frac{1}{2}\right)}
\int_{-\infty}^{\infty}d\theta\, P^{\mp}_\theta(t,x)\,e^{\mp\left(i\Omega-\frac{1}{2}\right)\theta}
\right]e^{i\vec{k}\cdot\vec{x}}. 
\end{equation}
As opposed to Eq.~\eqref{RindinMink}, these
functions are globally defined (and analytical) on the entire Minkowski
manifold, except for the origin. 
Furthermore, consistently with our previous considerations, 
they are eigenfunctions of the boost
generator, with eigenvalue $\Omega$. 
The coefficient $N^{\mp}_\kappa={(2\pi\sqrt{\mu_{\vec{k}}})}^{-1}\,e^{\pm \pi\Omega/2}$ is now fixed by requiring the orthonormality 
with respect to the ordinary inner product in Minkowski spacetime.
We also notice that the existence of two different
integral representations of the Bessel function
is reflected into the existence of two different
global representations of the Rindler modes
in Minkowski coordinates\footnote{The modes~\eqref{global} represent the spinorial counterpart of the Gerlach's Minkowski Bessel functions
for the scalar field~\cite{Gerlach,Luciano}.}. In turn, this 
is related to the possibility of extending such modes across the event horizons along two different paths (see~\cite{Oriti}
for more details).

At this stage, we can exploit the fact that the functions
$\psi^{+}_{j,\kappa}$ ($\psi^{-}_{j,\kappa}$) are 
linear combinations of only positive- (negative-) frequency
planewaves to introduce a field expansion
equivalent to Eq.~\eqref{eqn:planewaveexpansion} 
as regards the associated Fock space. 
This expansion reads
\be
\label{boostexp}
\Psi(t,\textbf{x})=\sum_{j=1,2}\int d^3\kappa\left[
c_{j,\kappa} \psi^-_{j,\kappa}(t,\textbf{x})+d^\dagger_{j,\kappa}\psi^+_{j,\kappa}(t,\textbf{x})
\right].
\ee
Since a field excitation in this quantization is still a positive-frequency
mode with respect to the time $t$, 
it is easy to understand that the vacuum
for $c_{k,\kappa}$ and $d_{k,\kappa}$
is the same as that for $a^r_{\hspace{0.2mm}\bk}$
and $b^r_{\hspace{0.2mm}\bk}$ defined in Eq.~\eqref{Minkvac}. 
Mathematically, this can be seen
by the following relations
\be
\label{eqn:c}
c_{j,\kappa}=\left(\psi^-_{j,\kappa}, \Psi\right)
= \sum_{r=1,2}\int{dk_x}\,F_{\hspace{0.2mm}j,r}(k_x,\Omega)\,a^r_{\hspace{0.2mm}\bk}\,,
\ee
\be
\label{F}
F_{\hspace{0.2mm}j,r}(k_x,\Omega)=\frac{4\pi^3}{\omega_{\textbf{k}}}\,N^-_{\kappa}\,N\,\left[{\left(\frac{\omega_{\textbf{k}}+k_x}{\omega_{\textbf{k}}-k_x}\right)}^{i\hspace{0.1mm}\frac{\Omega}{2}+\frac{1}{4}}e^{i\frac{\pi}{2}\left(i\Omega+\frac{1}{2}\right)}\hspace{0.9mm} X^\dagger_{j,\vec{k}}
+
{\left(\frac{\omega_{\textbf{k}}+k_x}{\omega_{\textbf{k}}-k_x}\right)}^{i\hspace{0.1mm}\frac{\Omega}{2}-\frac{1}{4}}e^{i\frac{\pi}{2}\left(i\Omega-\frac{1}{2}\right)}\hspace{0.9mm} Y^\dagger_{j,\vec{k}}\right]u^r_{\hspace{0.2mm}\bk}
\ee
which show that the $c_{j,\kappa}$-annihilator 
is a linear superposition of the $a^r_{\hspace{0.2mm}\bk}$'s only.
Similarly, for $d^\dagger_{j,\kappa}\psi^+_{j,\kappa}$ we have
\be
\label{eqn:d}
d^\dagger_{j,\kappa}=\left(\psi^+_{j,\kappa}, \Psi\right)
= \sum_{r=1,2}\int{dk_x}\,G_{\hspace{0.2mm}j,r}(k_x,\Omega)\,b^{r\dagger}_{\hspace{0.2mm}\bk}\,,
\ee
\be
\label{G}
G_{\hspace{0.2mm}j,r}(k_x,\Omega)=\frac{4\pi^3}{\omega_{\textbf{k}}}\,N^+_{\kappa}\,N\,\left[{\left(\frac{\omega_{\textbf{k}}+k_x}{\omega_{\textbf{k}}-k_x}\right)}^{i\hspace{0.1mm}\frac{\Omega}{2}+\frac{1}{4}}e^{i\frac{\pi}{2}\left(i\Omega+\frac{1}{2}\right)}\hspace{0.9mm} X^\dagger_{j,\vec{k}}
+
{\left(\frac{\omega_{\textbf{k}}+k_x}{\omega_{\textbf{k}}-k_x}\right)}^{i\hspace{0.1mm}\frac{\Omega}{2}-\frac{1}{4}}e^{i\frac{\pi}{2}\left(i\Omega-\frac{1}{2}\right)}\hspace{0.9mm} Y^\dagger_{j,\vec{k}}\right]v^r_{\hspace{0.2mm}\bk}
\ee
which show that the $d_{j,\kappa}$-annihilator 
is a linear superposition of the $b^r_{\hspace{0.2mm}\bk}$'s only.
To simplify the notation, in the above formulas
we have omitted the dependence of $F_{j,r}$
and $G_{j,r}$ by the transverse momentum $\vec{k}$, 
since it does not play any significant r\^ole in our next calculations. 
Furthermore, the field expansion~\eqref{eqn:planewaveexpansion}
has been used as right entry for the computation of the Minkowski inner products~\eqref{eqn:c} and~\eqref{eqn:d}. 

By using Eq.~\eqref{Minkvac}, it goes without saying that
\be
\label{equivquantconst}
c_{j,\kappa}|0\rangle_{\mathrm{M}}=d_{j,\kappa}|0\rangle_{\mathrm{M}}=0\,, \forall \kappa, j\,.
\ee 
The proof of the equivalence between
the two quantum constructions is completed by noticing that the transformations~\eqref{eqn:c} and~\eqref{eqn:d} are canonical, which implies that the new ladder operators $c_{j,\kappa}$ and $d_{j,\kappa}$ still obey the canonical anticommutation relations. 

\subsection{Unruh effect}
We can now derive the Unruh effect for the Dirac field
as originally proposed in~\cite{Unruh}. Such computation
requires the formulation of a quantum construction
which is valid for the Minkowski spacetime 
and gives back the quantization~\eqref{Rindfi}
when restricted to the $R$-wedge. 
For this task, we make use of the global modes
$\psi^{\mp}_{j,\kappa}$ and define the two combinations
\be
\mathcal{R}_{j,\kappa}\,=\,\frac{1}{\sqrt{2\cosh\left(\pi\hspace{0.2mm}\Omega\right)}}\left(e^{\frac{\pi\hspace{0.2mm}\Omega}{2}}\,\psi^-_{j,\kappa}\,+\,e^{-\frac{\pi\hspace{0.2mm}\Omega}{2}}\,\psi^+_{j,\kappa}\right),\qquad\, 
\mathcal{L}_{j,\kappa}\,=\,\frac{1}{\sqrt{2\cosh\left(\pi\hspace{0.2mm}\Omega\right)}}\left(e^{-\frac{\pi\hspace{0.2mm}\Omega}{2}}\,
\psi^-_{j,\kappa}\,-\,e^{\frac{\pi\hspace{0.2mm}\Omega}{2}}\,\psi^+_{j,\kappa}\right)\,.
\ee
These functions have all the required properties. 
Indeed, they are analytical and orthonormal in the whole Minkowski space. Furthermore $\mathcal{R}_{j,\kappa}$ is such that
it is defined everywhere but in the $L$-sector and reduces
to Eqs.~\eqref{eqn:psirindcoord} and~\eqref{eqn:psirindcoord2} in $R$ (depending on whether $j=1,2$).
The opposite behavior holds for $\mathcal{L}_{j,\kappa}$.

By inverting the above relations with respect
to $\psi^{\mp}_{j,\kappa}$ and replacing into Eq.~\eqref{boostexp}, 
we finally get
\begin{equation}
\label{eqn:RLexpan}
\psi(t,\bx)=\sum_{j=1,2}\int d^3\kappa\left[\hspace{0.2mm} r_{j,\kappa}\,\mathcal{R}_{j,\kappa}(t,\bx)\,+\,r^\dagger_{j,\kappa}\,\mathcal{R}_{j,\tilde\kappa}(t,\bx)\,+\,
l_{j,\kappa}\,\mathcal{L}_{j,\tilde\kappa}(t,\bx)\,+\,l^\dagger_{j,\kappa}
\,\mathcal{L}_{j,\kappa}(t,\bx)\right], 
\end{equation}
where the operators $r_{j,\kappa}$ and $l_{j,\kappa}$ are connected
to $c_{j,\kappa}$ and $d_{j,\kappa}$ via the Bogoliubov transformations
\be
\label{Bog}
r_{j,\kappa}=\frac{c_{j,\kappa}\,e^{\frac{\pi\hspace{0.2mm}\Omega}{2}}\,+\,d^\dagger_{j,\kappa}\,e^{-\frac{\pi\hspace{0.2mm}\Omega}{2}}}{\sqrt{2\cosh\hspace{0.2mm} (\pi\hspace{0.2mm}\Omega)}}\,,\qquad
l^\dagger_{j,\kappa}=\frac{c_{j,\kappa}\,e^{-\frac{\pi\hspace{0.2mm}\Omega}{2}}\,-\,d^\dagger_{j,\kappa}\,e^{\frac{\pi\hspace{0.2mm}\Omega}{2}}}{\sqrt{2\cosh\hspace{0.2mm} (\pi\hspace{0.2mm}\Omega)}}\,.
\ee
If we now identify $r_{j,\kappa}$ with 
the creation operator for the Rindler observer in $R$, we can
evaluate the spectrum of particles
detected by such an observer in the Minkowski vacuum quite
straightforwardly, obtaining
\begin{equation}
\label{eqn:thermal}
_\mathrm{M}\langle0|\hspace{0.2mm}r^\dagger_{j,\kappa}r_{i,\kappa'}|0\rangle_{\mathrm{M}}=\frac{1}{e^{2\pi\Omega}+1}\hspace{0.3mm}\delta_{ij}\,\delta^{\hspace{0.2mm}3}(\kappa-\kappa')\equiv\frac{1}{e^{\beta E}+1}\hspace{0.3mm}\delta_{ij}\,\delta^{\hspace{0.2mm}3}(\kappa-\kappa')\,,
\end{equation}
where in the last step we have introduced 
the Rindler proper energy $E=a\Omega$ and 
the inverse Unruh temperature $\beta=1/T_{\mathrm{U}}$.
Clearly, the same result would be obtained 
by considering the vacuum expectation value of
$l^\dagger_{j,\kappa}l_{i,\kappa'}$, which 
represents the particle number operator
for an accelerated observer in $L$. 
Thus, from Eq.~\eqref{eqn:thermal} we 
conclude that the Minkowski vacuum 
appears as a thermal state for the Rindler observer.
The spectrum of particle is distributed according
to the Fermi-Dirac statistics with temperature proportional to the magnitude of the acceleration. This is the well-known Unruh effect~\cite{Unruh}. 

A comment is in order here: the spectrum~\eqref{eqn:thermal}
diverges for $\kappa=\kappa'$. This is due to the fact
the states created by the action of $r^\dagger_{j,\kappa}$ 
and $a^{r,\dagger}_{\textbf{k}}$ on the vacuum are ill-defined, in the
sense that they are not properly normalizable. 
To avoid unphysical divergences, a wavepacket
based approach should be performed. 
This has been done explicitly for bosons in~\cite{qLuciano,Takagi}, 
where it has been shown that the ensuing spectrum 
retains the same profile as in the planewave formalism, 
with the Dirac delta function in energy-momentum
being now replaced by the Kronecker delta. 
However, since wavepackets do not affect the spinorial 
structure of the field, a similar result is expected to be
obtained for fermions as well. Then, in the wavepacket approach 
we can formally write 
\be
_\mathrm{M}\langle0|\hspace{0.2mm}r^\dagger_{j,\kappa}r_{i,\kappa'}|0\rangle_{\mathrm{M}}=\frac{1}{e^{2\pi\Omega}+1}\hspace{0.3mm}\delta_{ij}\hspace{0,2mm}\delta_{\kappa\kappa'}\,,
\ee
without any pathological behavior for fixed $\kappa=\kappa'$. 
In the next Section, 
we shall go on working with planewaves and only at
the end we take care of this subtlety.

\section{Flavor mixing in Rindler spacetime}
\label{QFTRIND}
Quantum mixing is among
the most challenging and important topics in Particle Physics.
In the well-established quantum mechanical version, it consists in the superposition of two states of particles with
different masses, rotated each other of an angle 
$\theta$ to give rise to the so-called flavor states. 
Since its prediction by Pontecorvo~\cite{Pont}, the theoretical basis
of flavor mixing, and in particular of neutrino mixing, 
have been largely investigated~\cite{GiuntiKim}, 
leading to the discovery of the phenomenon of neutrino
oscillations. Experimental developments have
later provided convincing evidences supporting
Pontecorvo's proposal~\cite{Experiments1,Experiments2}.
This has inevitably opened up new
scenarios in the physics beyond the Standard Model, 
which account for the neutrino having a non-vanishing
mass. 

Two decades later Pontecorvo's quantum mechanical approach, 
a field theoretical formalism for mixed fields
has been developed~\cite{BV95}. This has shown the
shortcomings of the original description
by pointing out the ortoghonality between the
vacuum for fields with definite flavors and that for fields with definite masses. Remarkably, this analysis has revealed 
a rich structure of the interacting-field vacuum
as a $su(2)$ coherent state, which in turn underlies
the unitary inequivalence between 
the flavor and mass Fock spaces and the 
consequent  alteration of the oscillation
formula to include the antiparticle degrees of freedom (see Appendix~\ref{Qmix} for more details). 

All of the above field theoretical considerations on flavor mixing
have been carried out in Minkowski background. 
Recently, they have been generalized to Rindler metric
for the case of boson mixing~\cite{Luciano}. 
To figure out how the flavor/mass inequivalence
appears to the Rindler observer, let us consider
the mixing transformations for Dirac fields in a simplified two-flavor model
\begin{subequations}
\label{Pontec}
\begin{equation}
\Psi_e(t,\textbf{x})\,=\,\cos\theta\,\Psi_1 (t,\textbf{x})+\sin\theta\,\Psi_2 (t,\textbf{x})\,,\\[-2.5mm]
\end{equation}
\begin{equation}
\hspace{3mm}\Psi_\mu (t,\textbf{x})\,=\,-\sin\theta\,\Psi_1(t,\textbf{x})+\cos\theta\,\Psi_2 (t,\textbf{x})\,, 
\end{equation}
\end{subequations}
where $\Psi_e$, $\Psi_\mu$ are the interacting fields with definite flavors (electron and muon), while $\Psi_1$, $\Psi_2$ denote the free fields with definite masses $m_1$ and $m_2$ (with $m_2>m_1$), satisfying the canonical 
anticommutator~\eqref{anticomcan}. 

By assuming the standard planewave expansion~\eqref{eqn:planewaveexpansion} for $\Psi_1$, $\Psi_2$, 
Eqs.~\eqref{Pontec} take the form~\cite{BV95}
\begin{equation}
\label{flavorfieldsexp}
\Psi_\ell(t,\textbf{x})=\sum_{r=1,2} \int d^{\hspace{0.2mm}3}\hspace{0.1mm}k\hspace{1mm} N \left[a^{r}_{\bk,\ell}(\theta,t)\hspace{0.7mm}u^{r}_{\bk,\sigma}(t)e^{i\bk\cdot\bx}\,+\,b^{r\hspace{0.2mm}\dagger}_{\bk,\ell}(\theta, t)\hspace{0.7mm}v^r_{\bk,\sigma}(t)e^{-i\bk\cdot\bx}\right],\quad (\ell,\sigma)=\{(e,1), (\mu,2)\},
\end{equation}
where $\ell$ ($\sigma$) is the flavor (mass) index and we have rewritten the expansion~\eqref{eqn:planewaveexpansion} in a formally different
(but equivalent) way by embedding the time-dependent exponential
factor in the spinors $u^{r}_{\bk,\sigma}(t)=u^r_{\bk,\sigma}e^{-i\omega_{\textbf{k},\sigma}t}$, 
$v^{r}_{\bk,\sigma}(t)=v^r_{\bk,\sigma}e^{i\omega_{\textbf{k},\sigma}t}$. The flavor annihilators $a^{r}_{\bk,\ell}(\theta,t)$ and $b^{r}_{\bk,\ell}(\theta,t)$ have been computed explicitly in Appendix~\ref{Qmix}. Furthermore, the algebraic nature of Eqs.~\eqref{Pontec}
has been explored in Eq.~\eqref{bogolgene}, showing
that each of them exhibits the structure
of a rotation nested into a Bogoliubov transformation
when considered at the level of ladder operators. 
As remarked above, the most striking implication of
this peculiar structure turns out to be the inequivalence between the mass vacuum $|0\rangle_{1,2}=|0\rangle_{1}\otimes|0\rangle_{2}$ (such that $a^r_{\bk,i}|0\rangle_{1,2}=b^r_{\bk,i}|0\rangle_{1,2}=0$, $i=1,2$) and the flavor vacuum $|0(\theta,t)\rangle_{e,\mu}$ (such that $a^r_{\bk,\ell}(\theta,t)|0(\theta,t)\rangle_{e,\mu}=b^r_{\bk,\ell}(\theta,t)|0(\theta,t)\rangle_{e,\mu}=0$, $\forall t$, $\ell=e,\mu$), 
which becomes a condensate of entangled massive particle/antiparticle pairs of density given by Eq.~\eqref{dens}. 

Now, in the previous Section we have seen 
that a useful tool to quantize the Dirac field
for an accelerated observer is the boost-mode expansion~\eqref{boostexp}.
Thus, to extend the QFT formalism of mixing 
to the Rindler metric,  we first consider
the analogue of Eq.~\eqref{boostexp} for flavor
fields. By using the same approach as that adopted
for the derivation of Eq.~\eqref{flavorfieldsexp}, we
infer the following relation
\be
\label{boostexpfl}
\Psi_\ell(t,\textbf{x})=\sum_{j=1,2}\int d^3\kappa\left[
c_{j,\kappa,\ell}(\theta,t)\, \psi^-_{j,\kappa,\sigma}(t,\textbf{x})+d^\dagger_{j,\kappa,\ell}(\theta,t)\,\psi^+_{j,\kappa,\sigma}(t,\textbf{x})
\right], \quad (\ell,\sigma)=\{(e,1), (\mu,2)\},
\ee
where the flavor ladder operators $c_{j,\kappa,\ell}$ and
$d_{j,\kappa,\ell}$ are related to the flavor 
operators $a^{r}_{\bk,\ell}$ and $b^{r}_{\bk,\ell}$ in the planewave expansion~\eqref{flavorfieldsexp} by
\be
\label{cflavor}
c_{j,\kappa,\ell}(\theta,t)
= \sum_{r=1,2}\int{dk_x}\,F_{\hspace{0.2mm}j,r,\sigma}(k_x,\Omega)\,a^r_{\hspace{0.2mm}\bk,\ell}(\theta,t)\,, \qquad (\ell,\sigma)=\{(e,1), (\mu,2)\},
\ee
with $F_{\hspace{0.2mm}j,r,\sigma}$ being given in Eq.~\eqref{F}, and
\be
\label{dflavor}
d^\dagger_{j,\kappa\,\ell}(\theta,t)
= \sum_{r=1,2}\int{dk_x}\,G_{\hspace{0.2mm}j,r,\sigma}(k_x,\Omega)\,b^{r\dagger}_{\hspace{0.2mm}\bk,\ell}(\theta,t)\,, \qquad (\ell,\sigma)=\{(e,1), (\mu,2)\},
\ee
with $G_{\hspace{0.2mm}j,r,\sigma}$ being given in Eq.~\eqref{G}
(see Eqs.~\eqref{eqn:c} and~\eqref{eqn:d} for the corresponding
relations in the mass basis). By relying on Eq.~\eqref{equivquantconst} and the related discussion, we stress that Eqs.~\eqref{cflavor} and~\eqref{dflavor} are canonical transformations
which leave the flavor vacuum $|0(\theta,t)\rangle_{e,\mu}$
unchanged. 

The expression~\eqref{boostexpfl} provides us with the 
springboard we were looking for. Indeed, we can closely 
follow the Rindler quantization scheme developed for massive fields  at the end of Sec.~\eqref{UE} and apply to flavor fields. 
In doing so, we arrive at the following expansion
\begin{equation}
\label{eqn:RLexpanfla}
\hspace{-1mm}\psi_\ell(t,\bx)=\sum_{j=1,2}\int d^3\kappa\left[\hspace{0.2mm} r_{j,\kappa,\ell}(\theta,t)\,\mathcal{R}_{j,\kappa,\sigma}(t,\bx)\,+\,r^\dagger_{j,\kappa,\ell}(\theta,t)\,\mathcal{R}_{j,\tilde\kappa,\sigma}(t,\bx)\,+\,
l_{j,\kappa,\ell}(\theta,t)\,\mathcal{L}_{j,\tilde\kappa,\sigma}(t,\bx)\,+\,l^\dagger_{j,\kappa,\ell}(\theta,t)
\,\mathcal{L}_{j,\kappa,\sigma}(t,\bx)\right], 
\end{equation}
where $(\ell,\sigma)=\{(e,1), (\mu,2)\}$, as usual. 
The flavor ladder operators for the Rindler observer
read
\be
\label{BogBog}
r_{j,\kappa,\ell}(\theta,t)=\frac{c_{j,\kappa,\ell}(\theta,t)\,e^{\frac{\pi\hspace{0.2mm}\Omega}{2}}\,+\,d^\dagger_{j,\kappa,\ell}(\theta,t)\,e^{-\frac{\pi\hspace{0.2mm}\Omega}{2}}}{\sqrt{2\cosh\hspace{0.2mm} (\pi\hspace{0.2mm}\Omega)}}\,,\qquad
l^\dagger_{j,\kappa,\ell}(\theta,t)=\frac{c_{j,\kappa,\ell}(\theta,t)\,e^{-\frac{\pi\hspace{0.2mm}\Omega}{2}}\,-\,d^\dagger_{j,\kappa,\ell}(\theta,t)\,e^{\frac{\pi\hspace{0.2mm}\Omega}{2}}}{\sqrt{2\cosh\hspace{0.2mm} (\pi\hspace{0.2mm}\Omega)}}\,,
\ee
which have indeed the same structure as the 
transformations~\eqref{Bog}, the mass index
being now replaced by the flavor one. 

Two comments deserve attention here: first, 
as opposed to the Minkowski frequencies $\omega_{\bk,i}$, $i=1,2$, 
the Rindler frequency $\Omega$ does not carry 
any index since it is independent of the field masses.
Furthermore, we emphasize that the flavor ladder operators for the Rindler observer are given by the combinations of two Bogoliubov
transformations, the thermal one responsible
for the Unruh effect and encoded by the $\Omega$-dependent
Bogoliubov coefficients, the second one
arising from mixing and implicitly contained in 
$c_{j,\kappa,\ell}$ and $d_{j,\kappa,\ell}$
trough Eqs.~\eqref{cflavor} and~\eqref{dflavor}.
This is the same structure found 
for the case of boson field mixing in Rindler space 
(see~\cite{Luciano,Blasone}). 

By using Eqs.~\eqref{eqn:annihilator} and~\eqref{annihilatorb}, 
we can now explicitly write down $r_{j,\kappa,\ell}$ as
\begin{eqnarray}
\non
r_{j,\kappa,\ell}(\theta,t)&=&
\frac{1}{\sqrt{2\cosh\hspace{0.2mm} (\pi\hspace{0.2mm}\Omega)}}\hspace{0.2mm}\,
{\sum_{r=1,2}\int{dk_x}\Big\{e^{\frac{\pi\hspace{0.2mm}\Omega}{2}}F_{\hspace{0.2mm}j,r,\sigma}(k_x,\Omega)\,a^r_{\hspace{0.2mm}\bk,\ell}(\theta,t)\,+\,}\,e^{-\frac{\pi\hspace{0.2mm}\Omega}{2}}G_{j,r,\sigma}(k_x,\Omega)\,\,b^{\hspace{0.2mm}r\hspace{0.2mm}\dagger}_{\bk,\ell}(\theta,t)\Big\}\,,\\[2mm]
\non
&&\hspace{-2.5cm}=\frac{1}{\sqrt{2\cosh\hspace{0.2mm} (\pi\hspace{0.2mm}\Omega)}}\hspace{0.2mm}\,
\sum_{r=1,2}\int{dk_x}\Bigg\{e^{\frac{\pi\hspace{0.2mm}\Omega}{2}}F_{\hspace{0.2mm}j,r,\sigma}(k_x,\Omega)\,\bigg\{
\cos\theta\,a^r_{\bk,1}+\sin\theta\sum_{s=1,2}\left[ (u^{r\dagger}_{\bk,1}(t),u^s_{\bk,2}(t))\, a^s_{\bk,2}+ (u^{r\dagger}_{\bk,1}(t),v^s_{-\bk,2}(t))\,b^{s\dagger}_{-\bk,2}\right]\bigg\}\\[2mm]
&&\hspace{-2.1cm}+\,e^{-\frac{\pi\hspace{0.2mm}\Omega}{2}}G_{\hspace{0.2mm}j,r,\sigma}(k_x,\Omega)\,\bigg\{
\cos\theta\,b^{r\,\dagger}_{\bk,1}\,+\,\sin\theta\sum_{s=1,2}\left[ (v^{s\dagger}_{\bk,2}(t),v^r_{\bk,1}(t))^*\, b^{s\dagger}_{\bk,2}\,+\, (u^{s\dagger}_{-\bk,2}(t),v^r_{\bk,1}(t))^*\,a^{s}_{-\bk,2}\right]\bigg\}\Bigg\}\,,
\end{eqnarray}
and similarly for $l_{j,\kappa,\ell}$. 

As in Eq.~\eqref{eqn:thermal}, we can
now derive the vacuum distribution
of mixed particles for the Rindler observer
by computing the expectation value
$_\mathrm{1,2}\langle0|\hspace{0.2mm}r^\dagger_{j,\kappa,\ell}(\theta,t)\,r_{j,\kappa,\ell}(\theta,t)|0\rangle_{\mathrm{1,2}}$.
However, this faces us with the computation of nontrivial integrals, 
which cannot be solved analytically. 
Despite such technicalities, 
interesting implications can still be derived 
by setting $t=\eta=0$ and in the realistic approximation
of relativistic neutrinos, i.e. $\epsilon\equiv\sqrt{m_2^2-m_1^2}/\omega_{\bk,1}\ll1$. Under these assumptions, 
it is a matter of direct calculations to show that
\begin{eqnarray}
\non
r_{j,\kappa,\ell}(\theta)|0\rangle_{\mathrm{1,2}}&=&
\frac{e^{\frac{\pi\Omega}{2}}}{\sqrt{2\cosh\hspace{0.2mm} (\pi\hspace{0.2mm}\Omega)}}\hspace{0.2mm}\,
\sum_{r=1,2}\int{dk_x}F_{\hspace{0.2mm}j,r,\sigma}(k_x,\Omega)\sin\theta\left(V_\bk^{r1}\,b^{1\dagger}_{-\bk,2}+V_\bk^{r2}\,b^{2\dagger}_{-\bk,2}
\right)|0\rangle_{\mathrm{1,2}}\\[2mm]
&&+\,\frac{e^{-\frac{\pi\Omega}{2}}}{\sqrt{2\cosh\hspace{0.2mm} (\pi\hspace{0.2mm}\Omega)}}\hspace{0.2mm}\,
\sum_{r=1,2}\int{dk_x}G_{\hspace{0.2mm}j,r,\sigma}(k_x,\Omega)
\left(\cos\theta\,b^{r\,\dagger}_{\bk,1}+\sin\theta\, U_\bk\,b^{r\,\dagger}_{\bk,2}
\right)|0\rangle_{\mathrm{1,2}}\,,
\end{eqnarray}
where the $U_\bk$ and $V_\bk$ coefficients
have been defined in Eqs.~\eqref{v11}-\eqref{uk}.
We stress that all the time-dependent quantities in the above equation must be intended as evaluated for $t=\eta=0$. 

Therefore, the particle spectrum at $t=\eta=0$ becomes
\begin{equation}
\label{moddi}
_\mathrm{1,2}\langle0|\hspace{0.2mm}r^\dagger_{j,\kappa,\ell}(\theta)\,r_{j,\kappa,\ell}(\theta)|0\rangle_{\mathrm{1,2}}=\frac{1}{e^{2\pi\Omega}+1}\hspace{0.3mm}\,+\,\frac{\sin^2\theta}{{2\cosh(\pi\Omega)}}\Big[e^{\pi\Omega}\, \mathcal{N}^{FF}_{j,\kappa,\sigma}\,-\,e^{-\pi\Omega}\, \mathcal{N}^{GG}_{j,\kappa,\sigma}\,+\,\big(\mathcal{N}^{FG}_{j,\kappa,\sigma}\,+\, \mathrm{c.c.}\big)
\Big]\,,
\ee
where the first contribution is the standard Fermi-Dirac 
distribution, while the terms multiplied by $\sin^2\theta$ 
are the mixing-induced corrections. Here, we have
used the relation~\eqref{upv} for the Bogoliubov
coefficients $U_\bk$ and $V_\bk$. Furthermore, 
we have implicitly taken into account the advantages of the
wavepacket approach, which allows us to get rid of 
any pathological divergency in the evaluation of the spectrum
for fixed momentum $\kappa$
(see the discussion at the end of Sec.~\ref{UE}).

The explicit expressions of $\mathcal{N}^{FF}_{j,\kappa,\sigma}$, 
$\mathcal{N}^{GG}_{j,\kappa,\sigma}$ and $\mathcal{N}^{FG}_{j,\kappa,\sigma}$
are rather awkward to exhibit. Here we provide
their integral expressions
\begin{eqnarray}
\label{nff}
\mathcal{N}^{FF}_{j,\kappa,\sigma}&\equiv&\sum_{r=1,2}\int dk_x |F_{j,r,\sigma}(k_x,\Omega)|^2\, |V_{\bk}|^2\,,\\[2mm]
\label{ngg}
\mathcal{N}^{GG}_{j,\kappa,\sigma}&\equiv&\sum_{r=1,2}\int dk_x |G_{j,r,\sigma}(k_x,\Omega)|^2\, |V_{\bk}|^2\,,\\[2mm]
\label{nfg}
\mathcal{N}^{FG}_{j,\kappa,\sigma}&\equiv&\sum_{r=1,2}\int dk_x \,F_{j,r,\sigma}(k_x,\Omega)\left[G_{j,1,\sigma}^{\hspace{0.1mm}*}(-k_x,\Omega)\,V^{r1}_\bk\,+\,G_{j,2,\sigma}^{\hspace{0.1mm}*}(-k_x,\Omega)\,V^{r2}_\bk
\right]U_\bk\,.
\end{eqnarray}
For concreteness, we consider the case of spin $j=1$ and flavor $\ell=e$ in Eq.~\eqref{moddi} (the same considerations hold true for $j=2$
and/or $\ell=\mu$). In turn, this amounts to setting the mass index $\sigma$ equal to one. Then, by replacing Eqs.~\eqref{F},~\eqref{G} and~\eqref{eqn:Vk2} into~\eqref{nff} and~\eqref{ngg} and rewriting 
\be
m_2=\sqrt{m_1^2+\epsilon^2\,\omega_{\bk,1}^2}\,,\qquad \omega_{\bk,2}=\omega_{\bk,1}\sqrt{1+\epsilon^2}\,,
\ee
one can show that
\be
\mathcal{N}^{FF}_{1,\kappa,1}\sim \mathcal{O}\left(\epsilon^4\right)\,,\qquad \mathcal{N}^{GG}_{1,\kappa,1}\sim \mathcal{O}\left(\epsilon^4\right)\,.
\ee
On the other hand, from Eq.~\eqref{nfg} it follows that
\be
\mathcal{N}^{FG}_{1,\kappa,1}\simeq\int dk_x\,\epsilon^2\, \mathcal{I}_\kappa(k_x)\,,
\ee
where the function $\mathcal{I}_\kappa(k_x)
$ is explicitly given in Appendix~\ref{Int}. By further manipulation, we finally arrive at
\be
\mathcal{N}^{FG}_{1,\kappa,1}\simeq \frac{|\Delta m^2|}{\mu_{\vec{k},1}^2}\,e^{-\pi\Omega}\,\mathcal{H}(\mu_{\vec{k},1})\,, 
\ee
where we have factorized out the dependence
on both $|\Delta m^2|=|m^2_2-m_1^2|$ and $\Omega$, and
$\mathcal{H}(\mu_{\vec{k},1})$ denotes the (dimensionless) $k_x$-integral 
of the residual function. Although such integral cannot be
evaluated analytically, numerical estimations show that it
assumes finite values and its real part is positive, at least
in the regime of interest (see the discussion below Eq.~\eqref{b4}).

Thus, to the leading order in $\epsilon$ we can neglect
the corrections $\mathcal{N}^{FG}_{1,\kappa,1}$
and $\mathcal{N}^{GG}_{1,\kappa,1}$  in Eq.~\eqref{moddi} and retain
only $\mathcal{N}^{FG}_{1,\kappa,1}$. In this way
the modified Unruh spectrum takes the form
\be
\label{tocomp}
_\mathrm{1,2}\langle0|\hspace{0.2mm}r^\dagger_{1,\kappa,e}(\theta)\,r_{1,\kappa,e}(\theta)|0\rangle_{\mathrm{1,2}}\,\simeq\,\frac{1}{e^{2\pi\Omega}+1}\hspace{0.3mm}\,+\,\frac{|\Delta m^2|}{\mu_{\vec{k},1}^2}\sin^2\theta\hspace{0.2mm}\operatorname{Re}\{\mathcal{H}(\mu_{\vec{k},1})\}\,\frac{e^{-\pi\Omega}}{{\cosh(\pi\Omega)}}\,, 
\ee
where $\operatorname{Re}\{z\}$ indicates the real part of $z$. 
Notice that this has the same structure as the
modified spectrum found for mixing of bosons~\cite{Luciano}. 
Furthermore, the mixing-induced correction
tends to $|\Delta m^2|/\mu_{\vec{k},1}^2 \sin^2\theta\operatorname{Re}\{\mathcal{H}(\mu_{\vec{k},1})\}\ll 1$ for $\Omega\rightarrow0$, 
while it vanishes for large $\Omega$. 

Within the framework of Boltzmann-Gibbs
thermodynamics, the result~\eqref{moddi} (or equivalently~\eqref{tocomp})
is a signature of the broken thermality
of Unruh radiation for mixed fields. Such an effect
has been discussed in more detail in~\cite{Luciano}
for the case of boson mixing. Regardless of the specific
nature of the fields, it can essentially be traced back
to the complex nature of the quantum vacuum
for mixed fields. Indeed, whilst in the absence
of mixing this state is made up of one type of 
pairwise correlated particles/antiparticles, 
for mixed fields it contains pairs of both equal
and different types (see the discussion in Appendix~\ref{Qmix}).
This gives rise to a nontrivial modification
of the Unruh distribution, which however is 
recovered for $\theta=0$ and/or
$m_2=m_1$, consistently
with the vanishing of mixing. 
We further elaborate on the physical meaning of Eq.~\eqref{tocomp} in the next Section, where
we revisit the above effect in the context of nonextensive
Tsallis theory.

\section{q-generalized Tsallis statistics for neutrino mixing}
\label{EDM}
In Boltzmann-Gibbs statistical mechanics, 
entropy is as a measure of the number of possible
microscopic states of a system in thermodynamic equilibrium, consistent with its macroscopic thermodynamic properties. This definition
is enclosed in the well-known formula
\be
\label{BGent}
S_{\mathrm{BG}}=-\sum_{i=1}^Wp_i\log p_i\,,
\ee
for a set of $W$ discrete microstates, 
where $p_i$ is the probability of the $i$-th microscopic
configuration with the condition $\sum_{i=1}^{W}p_i=1$. 
Clearly, if probabilities are all equal, this takes the 
form $S_{\mathrm{BG}}=\log W$. 

It is immediate to check that the entropy as defined
above is additive, which means that, given
two probabilistically independent systems $A$ and $B$ with 
entropies $S_{\mathrm{BG}}(A)$ and $S_{\mathrm{BG}}(B)$,  
the total entropy is $S_{\mathrm{BG}}(A+B)=S_{\mathrm{BG}}(A)+S_{\mathrm{BG}}(B)$. Intuitively, this reflects the fact that, in the absence of interactions, the number
of total states of a composite systems is simply given
by the sum of the states of its constituent parts. 

Although Boltzmann-Gibbs statistics is the correct way of 
approaching thermodynamic ergodic systems, 
it has been argued that strong correlations may be sometime responsible 
for the emergence of exotic features that
lye outside the domain of description of the standard theory. 
The most emblematic example of systems 
falling within this category are the gravitational systems, 
as already pointed out by Gibbs at the beginning of 
the last century. Hence, in such cases a generalization of Boltzmann-Gibbs
statistics and the related additive entropy is unavoidable. 

Along this line, in~\cite{Tsallis1,Tsallis2,Tsallis3,Tsallis4} it has
been shown that a proper generalization of
Eq.~\eqref{BGent} for complex systems exhibiting long-range
interactions and/or spacetime entanglement is given by the following
nonadditive entropy
\be
\label{TE}
S_q\,=\,\frac{1-\sum_{i=1}^Wp_i^q}{q-1}\,=\,\sum_{i=1}^Wp_i\log_q \frac{1}{p_i}\,, \,\,\quad q\in\mathbb{R^+}\,, 
\ee
where
\be
\log_q z \equiv \frac{z^{1-q}-1}{1-q}, \quad (\log_1 z=\log z)\,.
\ee
This is the so-called $q$-Tsallis entropy
and the framework built upon it is the
nonextensive (or Tsallis) statistical mechanics. 
We notice that the Boltzmann-Gibbs definition~\eqref{BGent} 
is a special case of $S_q$ for $q\rightarrow1$. 
Moreover, by considering again two 
independent systems such that $p_{ij}^{A+B}=p_{i}^Ap_{j}^B, 
\, \forall {(i,j)}$, Eq.~\eqref{TE} yields
\be
\label{sqab}
S_{q}(A+B)=S_q (A) + S_q (B) + (1-q) S_q (A) S_q(B)\,,
\ee
which shows that $S_{q}$ is superadditive or subadditive, depending
on whether $q<1$ or $q>1$. Since the $q$-entropic
index quantifies the departure from Boltzmann-Gibbs, 
it is typically named nonextensive Tsallis parameter.
As remarked above, the definition~\eqref{TE} works quite
well for strongly gravitating systems, and in particular for black holes~\cite{Tsallis3}, 
although in recent years encouraging results
have been obtained in the study of the cosmic background radiation~\cite{App7,App8}, low-dimensional dissipative systems~\cite{Tsallis4}, solar neutrinos~\cite{App11} and polymer chains~\cite{Polch}, among others. 

The $q$-generalized entropy~\eqref{TE} naturally
leads to nontrivial modifications of the standard
tools of Boltzmann-Gibbs theory. For instance, 
in~\cite{AbePla} it has been discussed how the removal
of the assumption of extensivity of entropy from statistical mechanics
affects the core thermodynamic relations, including the
zero-th law of thermodynamics and the concepts
of temperature and pressure. Similarly, in~\cite{Buyu,Buyu2,Buyu3,Buyu4,Chen}
Tsallis entropy has been used to derive
the following generalized average occupational numbers
\be
\label{modFD}
N_q(E)=\frac{1}{\left[1+(q-1)\beta\hspace{0.2mm}E\right]^{1/(q-1)}\pm1}\,,
\ee
which are obtained through the usual procedure of maximizing Tsallis entropy under the constraints of keeping constant the average internal energy and the average number of particles. Here the upper (lower) sign refers to the generalized 
Fermi-Dirac (Bose-Einstein)
distribution\footnote{With regard to the $q$-generalized Fermi-Dirac distribution, it is worth noting that apparent violations of Pauli's principle arising in the nonextensive quantum thermostatistics at finite temperature have been discussed in~\cite{VioPp} in connection with the possible existence of dark (\`a la Bacry) fermions. }. Furthermore $\beta=1/T$, where $T$ is the temperature of the system. 
We remark that the usual Fermi-Dirac and Bose-Einstein
distributions are recovered in the $q\rightarrow1$ limit.

To make physical sense, the distributions~\eqref{modFD}
must be non-negative and real-valued. This gives rise to the
following constraint
\be
\label{constraint}
\left\{\begin{array}{rcl}
&&\hspace{-4mm}0\le E\le \left[(1-q)\beta\right]^{-1}\,\,\,\,\,\,\,\,\,\ \mathrm{for}\,\,\, q<1,\\[3mm]
&&\hspace{-4mm}E\ge0\hspace{3cm}\mathrm{for}\,\,\, q>1\,.
\end{array}
\right.
\ee
Following~\cite{AppBE}, we also point out that
Eq.~\eqref{modFD} only provides a preliminary
but still useful approximation. Indeed, one cannot
derive the exact analytical expression of the generalized distribution 
for arbitrary values of $q$, due to the nonseparability of the partition function.   Nevertheless, 
the error committed turns out to be fairly negligible at very low temperatures
for systems with large numbers of particles, as it is for quantum fields
(see~\cite{AppBE} for more detailed numerical estimations). 
Accordingly, since the Unruh temperature is extremely small 
even for huge accelerations\footnote{Let us remind that $T_{\mathrm{U}}$ 
is smaller than $1\,\mathrm{K}$ even for accelerations up to $10^{21}\,\mathrm{m/s^2}$.}, 
the use of Eq.~\eqref{modFD} is well justified.

In the previous Section we have emphasized
that the Unruh spectrum for mixed fields loses its typical thermal 
profile owing to the entangled structure acquired by the quantum vacuum, 
which becomes a (time-dependent) $su(2)$ coherent state
(see Eqs.~\eqref{moddi} or~\eqref{tocomp}). 
Based on the observation that the generalized Tsallis framework well describes
strongly correlated systems, either on quantum or classical grounds,
in~\cite{qLuciano} the question arose of how
such a result could appear when rephrased in the language
of nonextensive Tsallis statistics. Remarkably, for mixing of bosons  
it has been shown that the modified Unruh distribution can 
be mapped into the $q$-generalized Bose-Einstein distribution~\eqref{modFD}, provided that the $q$-entropic
index is properly related to the characteristic mixing parameters
$\sin\theta$ and $\Delta m$. 

Following this analysis, 
here we extend the Tsallis paradigm to mixing of fermions as well. 
Clearly, since the correction in Eq.~\eqref{tocomp} slightly affects the Fermi-Dirac
spectrum at both high and low energy regimes, 
it is reasonable to expand the generalized 
distribution~\eqref{modFD} (with the $+$ sign in the denominator)
for tiny departures 
of $q$ from unity. To the leading order, we then obtain
\begin{eqnarray}
\label{qmeno1}
\non
N_q(E)&\simeq&\frac{1}{e^{\beta E}+1}\,+\,\frac{(\beta E)^2}{4\left[1+\cosh(\beta E)\right]}(q-1)\,,\\[2mm]
&=&\frac{1}{e^{\beta E}+1}\,+\,\frac{1}{8}(\beta E)^2\,\mathrm{sech}^2\left(\frac{\beta E}{2}\right)(q-1)\,,
\end{eqnarray}
to be compared with the corresponding Tsallis-induced correction for bosons (see Eq.~(39)  of~\cite{qLuciano}) 
\be
\label{Loq}
N_q(E)\simeq\frac{1}{e^{\beta E}-1}\,+\,\frac{1}{8}\left(\beta E\right)^2\,\csch^2\left(\frac{\beta E}{2}\right)\left(q-1\right).
\ee

In passing, we note that a similar perturbative analysis
in nonextensive statistics has been carried out in 
the context of high-energy heavy-ion collisions~\cite{Alberico}, 
starting from the experimental evidence that high-energy nucleus-nucleus 
collisions cannot be described in terms of superpositions of elementary nucleon-nucleon interactions. In that case, the use of Tsallis statistics
is motivated by the hypothesis that collective effects in the hadronic medium
can be traced back to some kind of memory effects and long-range forces that
may occur during high-energy ion collisions. 

Now, Eq.~\eqref{qmeno1} can be compared to the modified spectrum~\eqref{tocomp}
by setting $E=a\Omega$ (the proper energy of Rindler modes) and
$\beta=1/T_{\mathrm{U}}=2\pi/a$. Straightforward calculations lead to
\be
\label{Tsfirstorder}
N_q(\Omega)\simeq\frac{1}{e^{2\pi\Omega}+1}\,+\,
\frac{\pi^2}{2}\,\Omega^2\,\mathrm{\sech}^2\left(\pi\hspace{0.3mm}\Omega\right)\left(q-1\right). 
\ee
Although we are only retaining the first
term in the expansion, we notice that 
the energy dependence of the $q$-generalized
spectrum gets nontrivially altered. 

Let us now investigate the correspondence between 
the modification of Unruh distribution 
arising in the context of flavor mixing 
and that obtained in the nonextensive Tsallis theory, respectively.
By equating Eqs.~\eqref{tocomp} and~\eqref{Tsfirstorder}, we infer
\be
q=1+\mathcal{F}_\theta(\Delta m^2,\mu_{\vec{k},1})\,\frac{1+e^{-2\pi\Omega}}{\Omega^2}\simeq 1+\frac{\mathcal{F}_\theta(\Delta m^2,\mu_{\vec{k},1})}{\Omega^2}\,,
\label{q1}
\ee
where in the last step we have approximated 
\be
\frac{1+e^{-2\pi\Omega}}{\Omega^2}\simeq \frac{1}{\Omega^2}, \qquad \mathrm{for}\,\,\Omega>0\,,
\ee 
apart from an irrelevant factor, and we have defined
\be
\mathcal{F}_\theta(\Delta m^2,\mu_{\vec{k},1})\equiv\frac{|\Delta m^2|\,\operatorname{Re}\{\mathcal{H}(\mu_{\vec{k},1})\}}{\pi^2\mu_{k,1}^{\hspace{0.1mm}2}\hspace{0.1mm}}\sin^2\theta\,.
\ee
The function $\mathcal{H}(\mu_{\vec{k},1})$ is defined in Eqs.~\eqref{b2}-\eqref{b4}
of Appendix~\ref{Int}. 

Now, since $\operatorname{Re}\{\mathcal{H}(\mu_{\vec{k},1})\}>0$
(at least in the regime we are considering, see Appendix~\ref{Int}), 
from the above relations we infer 
that $q>1$, indicating the subadditive regime of 
Tsallis entropy, according to Eq.~\eqref{sqab}. 
Furthermore, from Eq.~\eqref{constraint} we see
that there is no upper limit on the energy scale. 
By contrast, for the case of boson mixing one has
$q<1$, which corresponds to a flavor vacuum entropy
higher than the sum of the entropies
of the two mass vacua taken separately (Tsallis superadditive regime)~\cite{qLuciano}.

Therefore, in the above effective picture,
the properties of 
Unruh condensate for mixed particles can
be described in terms of the nonextensive Tsallis statistics, 
with the $q$-entropic index satisfying the condition~\eqref{q1}. 
As expected, the departure of $q$
from unity depends on the mixing angle 
and the mass difference in such a way
that the usual Boltzmann-Gibbs theory with 
$q=1$ is recovered for $\theta=0$ and/or $\Delta m=0$, 
consistently with the vanishing of mixing in both cases. 
The same behavior occurs in the ultrarelativistic limit 
$\Delta m/\mu_{\vec{k},1}\rightarrow0$, since $\operatorname{Re}\{\mathcal{H}(\mu_{\vec{k},1})\}$ is finite even for large momenta. This
can be understood by observing that in such approximation the field
theoretical description of flavor mixing reproduces the
standard quantum mechanical framework, in which
the flavor and mass representations are connected
by a simple rotation, and thus the related Fock spaces
are unitarily equivalent to each other (see the discussion in Appendix~\ref{Qmix}).

Following~\cite{qLuciano}, we now exploit Eq.~\eqref{q1}
to provide an alternative interpretation for the result~\eqref{tocomp}. 
Indeed, instead of regarding
the mixing-induced distortion of the spectrum 
as a nonthermal effect within Boltzmann-Gibbs framework, 
we can trace its origin back to a modification of the vacuum distribution of  
particles/antiparticle pairs at a more fundamental statistical level. 
We remark that a similar connection between
the Unruh effect and Tsallis statistics has been 
exhibited in~\cite{Shababi} in the context of deformed uncertainty relations accounting
for a minimal length at Planck scale (Generalized Uncertainty Principle).

As for the case of bosons, the identification in Eq.~\eqref{q1}
shows that the $q$-entropic index exhibits a running
behavior as a function of the Rindler frequency $\Omega$. 
This is quite a common behavior for quantum field theoretical
and quantum gravity systems. An analogous study in Tsallis statistics
with a varying nonextensive parameter has been recently 
developed in cosmology in~\cite{App13}. Interestingly, 
we notice that, while for mixing of bosons the deviation of $q$ from unity 
depends on $\Omega^{-1}$~\cite{qLuciano}, in the present case 
it scales as $\Omega^{-2}$. Of course, this does not entail any dimensional problem,
being the Rindler frequency $\Omega$ dimensionless. 

In this connection, one may argue that Eq.~\eqref{q1} exhibits a pathological behavior
for arbitrarily small frequencies, as it diverges in the $\Omega\rightarrow0$ limit. 
However, as discussed in~\cite{qLuciano}, this
must be interpreted as a warning concerning
the domain of validity of our approximation. 
Indeed, for $\Omega$ below a certain threshold $\Omega_{\min}$, 
it happens that the $q$-entropic index might
largely deviate from unity, which \emph{a posteriori}
would invalidate the expansion~\eqref{Tsfirstorder}.
To keep the self-consistency of our analysis, we have
to require that $|q-1|\ll1$. In turn, this gives rise
to the infrared frequency cutoff 
\be
\Omega\gg\Omega_{\min}=[F_\theta(\Delta m^2,\mu_{\vec{k},1})]^{1/2}=
\left[\frac{|\Delta m^2|\,\operatorname{Re}\{\mathcal{H}(\mu_{\vec{k},1})\}}{\pi^2\mu_{k,1}^{\hspace{0.1mm}2}\hspace{0.1mm}}\sin^2\theta\right]^{1/2}\,,
\ee 
which means that the more accurate the approximation of relativistic neutrinos $\Delta m/\mu_{\vec{k},1}\ll 1$, 
the higher the number of Rindler modes that fit with the $q$-generalized Fermi-Dirac  distribution~\eqref{modFD}. Clearly, the entire frequency spectrum
is spanned for $F_\theta\rightarrow0$. 
To give some numerical estimations, 
here we notice that, by setting the sampling values $\theta=\pi/4$, 
$\Delta m^2\sim10^{-5}\,\mathrm{eV}^2$ and $k_y=k_z\simeq 10\,\mathrm{GeV}$, 
we get $\Omega_{\min}\simeq10^{-8}$. 
On the other hand, the departure from 
extensivity becomes increasingly negligible for large $\Omega$. In particular
the Boltzmann-Gibbs statistics with $q=1$ is recovered for $\Omega\rightarrow\infty$.
As explained in~\cite{qLuciano}, this is somehow consistent with the
fact that the higher the energy of the level we are considering, the lower the expected
number of particles it can accommodate, with both the Fermi-Dirac and $q$-generalized distributions approaching zero as $\Omega\rightarrow\infty$. Therefore, 
it is reasonable to expect that the difference between 
the two spectra narrows for relatively large $\Omega$.
This is exactly the same behavior found in~\cite{qLuciano} for mixing of bosons.

 \begin{figure}[t]
\resizebox{5.5cm}{!}{\includegraphics{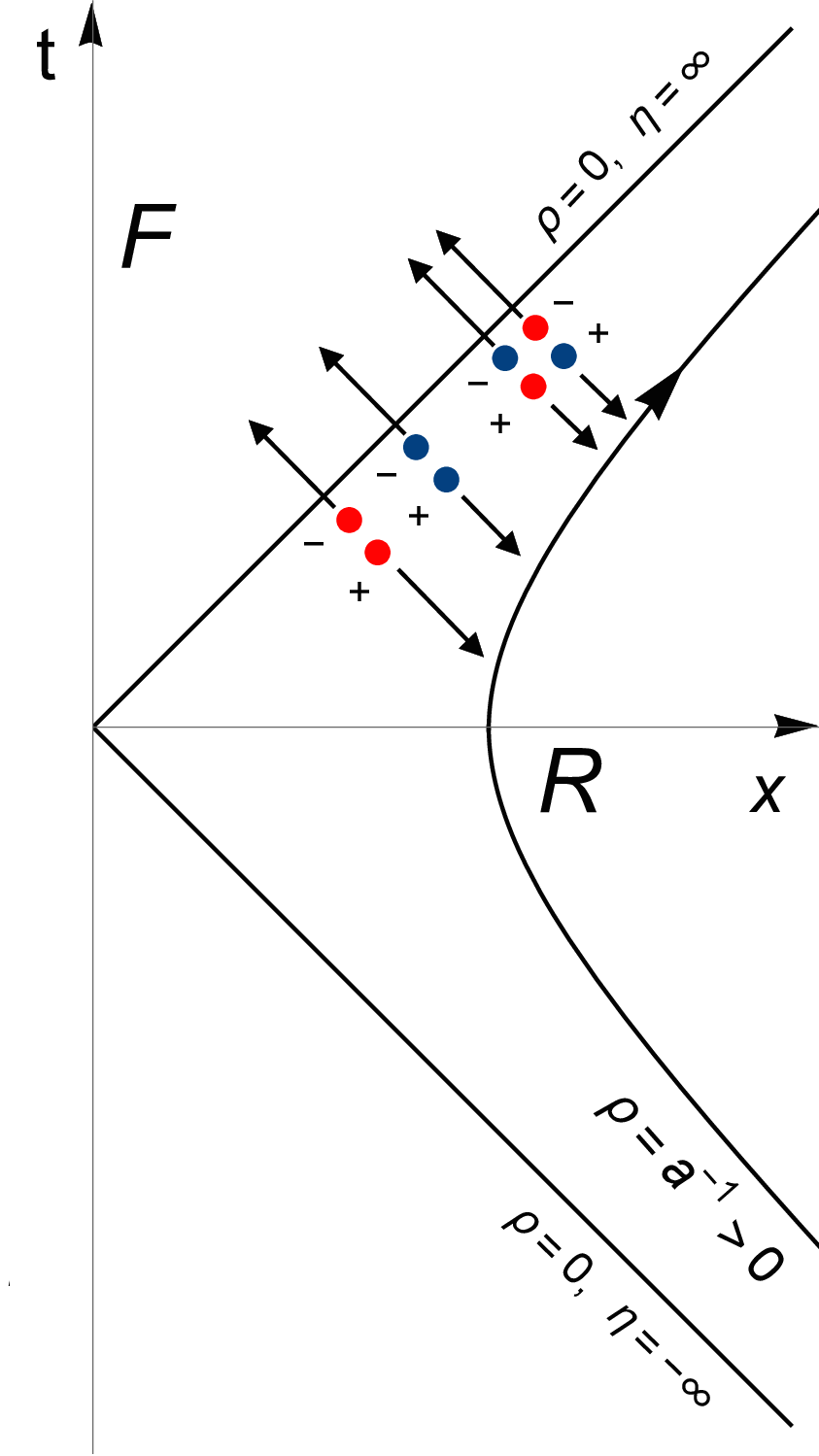}}
\caption{\small{Pictorial interpretation of the mixing-induced modification of Unruh condensate. Different (online) dot colors correspond to different types of pairwise correlated particles/antiparticles. Unruh effect originates from
vacuum fluctuations close to Rindler horizon.
For free (i.e unmixed) fields, the vacuum state is made up of single-type pairs (either blue-blue or red-red). By contrast, 
in the presence of field mixing 
its structure becomes far richer due to 
the simultaneous appearance of hybrid pairs (red-blue and blue-red). 
The ensuing spectrum can still be featured as a thermal-like bath, 
provided that vacuum constituents are assumed to be distributed 
according to nonextensive Tsallis statistics.}}
\label{Rindlerbis}
\end{figure}

The connection established between 
the modification~\eqref{tocomp} induced by flavor mixing
and the $q$-generalized Fermi-Dirac distribution~\eqref{qmeno1}
can be explained in terms of the nontrivial 
structure acquired by the vacuum for mixed fields (see Appendix~\ref{Qmix}). 
Indeed, due to the peculiar features of mixing transformations
in quantum field theory, this state turns out to be
a condensate of entangled particle/antiparticle 
pairs of both equal and different types.
As a consequence, while the standard Unruh effect
may be seen as due to single-type fluctuations spontaneously appearing
close to the Rindler horizon (one component falling back into the horizon, 
the other escaping as Unruh radiation), 
in the presence of field mixing it can be generated by
different types of virtual pairs, resulting in a 
deformed spectrum of Unruh radiation.
This has been pictorially represented in Fig.~\ref{Rindlerbis}
by associating different (online) colors to different
types of particles. We have shown above that such
effect can be framed in the context of Tsallis statistics, 
with a modified distribution given by the
$q$-generalized distribution~\eqref{qmeno1} based on
nonadditive Tsallis entropy~\eqref{TE}. 
In particular, the departure $q-1$ from extensivity 
is quantified by the characteristic mixing parameters and
the energy scale (see Eq.~\eqref{q1}). 
As a remark, from Eq.~\eqref{sqab} we also emphasize that $q>1$ implies
that mixed particles are packaged in the Unruh condensate
in such a way that the total entropy of the 
vacuum for mixed fields is lower than the sum
of the vacuum entropies for the corresponding free fields. 
Heuristically speaking, this can be understood as follows: suppose
we dispose of a detector of red-type particles only (analogous considerations
hold for a detector of blue-type particles). In the absence of mixing, 
the detection of a particle definitely corresponds to the same type of antiparticle
fallen back into the horizon. By contrast, for mixed fields, if we see
a red particle, it may be associated to a red-antiparticle, as well as to a blue-antiparticle  crossing inward the horizon, with
the sum of probabilities being normalized to unity (see Fig.~\ref{Rindlerbis}). 
However, a blue-antiparticle
that falls into the horizon can also lead to the escape of a blue-particle, 
which will remain undetected under our hypothesis.  
Since, for the Pauli exclusion principle, two identical fermions
cannot occupy the same quantum state simultaneously, we infer that 
the existence of blue-antiparticle/blue particle pairs in the vacuum
affects the microscopic configuration of the system
by reducing the available phase space volume for the blue-antiparticle/red particle pairs. Indeed, it may happen that the blue antiparticle of a blue-blue pair with given quantum numbers crosses inward the horizon
before the blue antiparticle of a blue-red pair with the same quantum numbers does.
Then, for its part the blue antiparticle of the blue-red pair will be prevented from
entering the horizon and its companion from escaping as Unruh radiation.
This results in a lower number of possible microstates for the observed
system and a related decrease of the total entropy with respect to the case of unmixed fields.
Of course, such a behavior does not occur for bosons, 
for which the condensate structure of vacuum made up 
of hybrid types of pairs, combined with the fact that
bosons are not subject to the Pauli exclusion principle, 
leads to the opposite superadditive regime for
Tsallis entropy~\cite{qLuciano}.

It is interesting to observe that a similar comparative
analysis of the generalized thermostatistic properties
of bosons and fermions has been carried out in~\cite{Lavagno}
in the context of the quantum $q$-deformed algebra.
This differs from Tsallis framework in that 
it relies on the modification of the quantum algebra of the creation and annihilation operators based on the $q$-calculus~\cite{Jack}, rather than the modification
of the partitioning of the microstates 
of the many body system. In that case, it has been shown
that $q$-deformed bosons and fermions have an
enhancement of the quantum statistical effects compared to standard behavior.
In addition, in~\cite{Mirza} the thermodynamic geometry of an ideal q-deformed boson and fermion gas has been constructed, investigating
some properties such as the stability
and statistical interaction. 

The above considerations allow us to state that 
the correlations induced by mixing do spoil
the macroscopic properties of Unruh 
condensate by affecting the 
statistical behavior of its microscopic configurations. 
In this connection, we notice that nonextensive
statistics based on Tsallis entropies, such as the relative entropy and the 
Peres criterion, have been widely used 
in the study of entanglement in the last years~\cite{TSENT}.

\section{Conclusions and Outlook} 
\label{Conc}
The Unruh effect is a notorious prediction 
built in quantum field theory, which states
that empty space appears as a thermal bath
of particles from the point of view of an accelerating
observer~\cite{Unruh}.
Recently, it has been shown that this effect
is nontrivially spoilt when dealing with mixing
of fields with different masses~\cite{Luciano,NonTN}, 
leading to a nonthermal distortion of the vacuum condensate.
Following the analysis of~\cite{qLuciano} for boson mixing, here we have
revisited this phenomenon in the context of nonextensive
Tsallis thermodynamics. By considering the 
superposition of two Dirac neutrino fields in the relativistic approximation, 
we have provided an effective description of 
the modified Unruh spectrum in terms of the 
$q$-generalized Fermi-Dirac distribution
based on the nonadditive Tsallis
entropy. In this framework, the departure
from the standard Boltzmann-Gibbs theory
turns out to be quantified by the mass difference of the mixed
fields and the mixing angle, in such a way that
the ordinary extensive mechanical statistics is recovered
for vanishing mixing. Furthermore, the condition $q>1$ 
indicates the subadditive feature of Tsallis entropy. 
In this sense, the present result differs from that of~\cite{qLuciano}, 
where the opposite superadditive regime has been obtained.

As remarked in~\cite{qLuciano}, the above analysis  
provides us with a more conservative interpretation
for the result of~\cite{Luciano,NonTN}, since it allows us 
to extend the usual thermal behavior of Unruh effect
to the case of mixed fields. In other terms, one can 
maintain a thermal-like picture for the vacuum condensate, 
provided that the distribution of particle/antiparticle
pairs is assumed to obey Tsallis (rather than Boltzmann) prescription.  
In turn, the origin of this statistical effect can be ascribed to the
nontrivial structure acquired by the vacuum for mixed fields, 
which becomes a hybrid condensate of
entangled particle-antiparticle pairs.  
Therefore, in this scenario it is still possible to associate
an effective temperature to the vacuum condensate, which would however be 
different from the standard definition~\eqref{TU}, 
since it would depend on the entropic index (see~\cite{AbePla} for more details). Clearly, this aspect deserves careful attention
and will be investigated in more detail elsewhere, 
along with the possibility of having a non-stationary
Unruh condensate in the presence of field mixing. 
In this respect, it should be emphasized that 
a $q$-dependent expression for the Unruh temperature 
has been derived in the context
of the Generalized Uncertainty Principle (GUP)
in~\cite{Shababi}. In that case, the departure of Unruh effect 
from Boltzmann statistics is induced by the emergence
of a minimal length of the order of Planck scale, 
which affects the phase-space structure by
changing the elementary cell volume. In light of this similarity, 
it would be interesting to investigate whether
there exists any kind of connection between 
the two frameworks. For instance, a suggestive idea could be
to rethink the phenomenon of mixing as an effect induced by
background geometry, just like the GUP
arises from a modified geometry of spacetime
at quantum gravity scale. 

Apart from these long-term perspectives, 
the present analysis can be refined by 
looking at the following issues: to avoid technicalities, 
in our computations we have considered
a simplified model involving only two fields.
However, we expect that the generalization
to the more realistic three-flavor scenario 
does not affect our results at the conceptual level.
Furthermore, we have made use of the assumption of relativistic neutrinos
and, at the same time, we have expanded
the $q$-generalized Fermi-Dirac distribution
for small deviations of $q$ from unity. To strengthen our outcome,  
the exact calculation would be required. 
Then, the question arises as to 
how Eq.~\eqref{q1} appears when relaxing any 
approximation. We also developed computations
for $t=0$, neglecting the time-dependence
of the vacuum for mixed fields. Clearly, a more complete
analysis should take into account the time evolution
of the vacuum (and any possible deviation from thermal equilibrium) as well. 
A further direction would be to
analyze the above framework by considering neutrinos as Majorana fields. 
As well-known, indeed, the very
nature of neutrinos -- Dirac or Majorana -- is still 
unclear, the only practical way to discriminate
being the search for neutrinoless double $\beta$-decays. 
Based on~\cite{Majorana}, where the Unruh effect
was shown to be insensitive to the 
Majorana/Dirac dichotomy, also in this case
we envisage no significant 
deviations from the current result.

Finally, it is easy to understand that experimental tests
of our prediction are extremely challenging at the present,
the reason being that Unruh radiation itself has not
yet been detected. Nevertheless, promising proposals
have been put forward in the context of analog models.
Among these, we mention experiments on water waves~\cite{Water,Water2},
graphene~\cite{Analog2} and other condensed matter systems~\cite{Analog,Analog3}.
Therefore, such models represent the only source
to address experimentally the Unruh effect -- and any possible deviation from the standard behavior -- to date.

\acknowledgements 

One of the authors (GGL) is grateful to Costantino
Tsallis (Centro Brasileiro de Pesquisas Fisicas, Brazil) for helpful conversations.

\appendix
\section{QFT of flavor mixing}
\label{Qmix}
This Appendix contains technical details on the
quantization of mixed Dirac fields in Minkowski spacetime~\cite{BV95}.
To avoid technicalities and make the physical insight
as transparent as possible, we focus on a toy model
involving two fields only. However, the same considerations
hold in the three-flavor case~\cite{3f}. 

To better understand the origin of the
genuinely QFT features of flavor mixing, 
let us rewrite the transformations~\eqref{Pontec}
in terms of the algebraic generator $G_\theta(t)$ as~\cite{BV95}
\begin{equation}
\label{bogolgene}
{\Psi}_\ell(t,\textbf{x})\,=\,{G}_\theta^{-1}(t)\,{\Psi}_\sigma(t,\textbf{x})\,{G_\theta}(t)\,,
\end{equation}
where $(\ell,\sigma)=\{(e,1),(\mu,2)\}$ and
\be
\label{Gene}
G_\theta(t)\,=\,\exp\lf\{\theta\int d^3x\,\lf[{\Psi}^\dagger_1(t,\textbf{x}){\Psi}_2(t,\textbf{x})-{\Psi}^\dagger_2(t,\textbf{x}){\Psi}_1(t,\textbf{x})\ri]\ri\}. 
\ee
By introducing the operators
\be
S_+\equiv \int d^3x\,\Psi^\dagger_1(t,\textbf{x})\Psi_2(t,\textbf{x})\,,\qquad S_-=(S_+)^\dagger\,,
\ee
$G_\theta$ can be cast in the equivalent form
\be
G_\theta(t)=\exp\left[\theta\left(S_+-S_-\right)\right].
\ee
It is easy to verify that $G_\theta(t)$ belongs to the $su(2)$
group, whose algebra is closed by
\begin{eqnarray}
S_3&=&\frac{1}{2}\int d^3x\left(\Psi_1^\dagger(t,\textbf{x})\Psi_1(t,\textbf{x})-\Psi_2^\dagger(t,\textbf{x})\Psi_2(t,\textbf{x}) \right)\,,\\[2mm]
S_0&=&\frac{1}{2}\int d^3x\left(\Psi_1^\dagger(t,\textbf{x})\Psi_1(t,\textbf{x})+\Psi_2^\dagger(t,\textbf{x})\Psi_2(t,\textbf{x}) \right).
\end{eqnarray}

In the QFT treatment of flavor mixing, the generator $G_\theta(t)$
provides the dynamical map between the Fock space $\mathcal{H}_{e,\mu}$ for definite flavor fields and the Fock space $\mathcal{H}_{1,2}$ for definite mass fields. We shall refer to these spaces
as flavor and mass Fock spaces, respectively~\cite{BV95}. Thus, we can write 
\be
\label{vac}
G_\theta^{-1}(t): \mathcal{H}_{1,2} \mapsto {\mathcal H}_{e,\mu}\,,
\ee
Specifically, for the mass vacuum $|0\rangle_{1,2}=|0\rangle_{1}\otimes|0\rangle_{2}$ the above relations leads to
\be
|0(\theta,t)\rangle_{e,\mu}=G_\theta^{-1}(t)|0\rangle_{1,2}\,,
\ee
where we have denoted by $|0(\theta,t)\rangle_{e,\mu}$ the 
vacuum for flavor fields. 

The peculiar properties of QFT mixing 
must be sought in the nontrivial nature of $G_\theta$.
Indeed, while for quantum mechanical
systems this is a unitary operator preserving
the canonical anticommutation relations, in QFT
it turns out to be nonunitary in the infinite volume limit.
From Eq.~\eqref{vac}, it follows that the vacua
$|0\rangle_{1,2}$ and $|0(\theta,t)\rangle_{e,\mu}$
become orthogonal to each other, resulting
in unitarily (i.e. physically) inequivalent mass and flavor Fock spaces.
Clearly, such inequivalence disappears for $\theta=0$ and/or
$m_2=m_1$, as expected in the absence of field mixing.

To figure out the effects of the dynamical map~\eqref{vac} on the
flavor vacuum, let us focus on the second quantization of the flavor
fields $\Psi_{e,\mu}$. As shown in Sec.~\ref{QFTRIND}, in the planewave representation the transformations~\eqref{bogolgene} yield the free-field like expansion~\eqref{flavorfieldsexp}, here rewritten for matter of convenience
\begin{equation}
\label{flavorfields}
\Psi_\ell(t,\textbf{x})=\sum_{r=1,2} \int d^{\hspace{0.2mm}3}\hspace{0.1mm}k\hspace{1mm} N \left[a^{r}_{\bk,\ell}(\theta,t)\hspace{0.7mm}u^{r}_{\bk,\sigma}(t)e^{i\bk\cdot\bx}\,+\,b^{r\hspace{0.2mm}\dagger}_{\bk,\ell}(\theta, t)\hspace{0.7mm}v^r_{\bk,\sigma}(t)e^{-i\bk\cdot\bx}\right],\quad (\ell,\sigma)=\{(e,1), (\mu,2)\},
\end{equation}
where the notation has been set in Sec.~\ref{QFTRIND}. 
The time-dependent flavor annihilators are defined by~\cite{BV95}
\be
\label{aflav}
a^{r}_{\bk,\ell}(\theta,t)=G^{-1}_\theta(t)\hspace{0.2mm}a^{r}_{\bk,\sigma}\hspace{0.2mm}G_\theta(t), \qquad (\ell,\sigma)=\{(e,1), (\mu,2)\},
\ee
and similarly for $b^{r}_{\bk,\ell}(\theta,t)$. To streamline
the notation, henceforth we shall denote these operators
by $a^{r}_{\bk,\ell}(\theta,t)\equiv a^{r}_{\bk,\ell}$. 
By computing the above product explicitly, we get
\begin{eqnarray}
\label{eqn:annihilator} 
a^r_{\bk,e}&=&\cos\theta\,a^r_{\bk,1}\,+\,\sin\theta\sum_{s=1,2}\left[ (u^{r\dagger}_{\bk,1}(t),u^s_{\bk,2}(t))\, a^s_{\bk,2}\,+\, (u^{r\dagger}_{\bk,1}(t),v^s_{-\bk,2}(t))\,b^{s\dagger}_{-\bk,2}\right],\\[2mm]
b^r_{\bk,e}&=&\cos\theta\,b^r_{\bk,1}\,+\,\sin\theta\sum_{s=1,2}\left[ (v^{s\dagger}_{\bk,2}(t),v^r_{\bk,1}(t))\, b^s_{\bk,2}\,+\, (u^{s\dagger}_{-\bk,2}(t),v^r_{\bk,1}(t))\,a^{s\dagger}_{-\bk,2}
\label{annihilatorb}
\right],
\end{eqnarray}
(and similarly for $a^r_{\bk,\mu}$ and $b^r_{\bk,\mu}$). 
Therefore, we find that the flavor annihilators 
are related to the ladders operators in the mass representations
via the combinations of two transformations: the standard Pontecorvo
rotation (parameterized by the mixing angle $\theta$) and
a Bogoliubov transformation (the terms in the square brackets) of coefficients\footnote{Here we only display the coefficients that explicitly appear in the computations of Sec.~\ref{QFTRIND}.}
\begin{eqnarray}
\label{v11}
V^{11}_\textbf{k}(t)&\equiv&(u^{1\dagger}_{\bk,1}(t),v^1_{-\bk,2}(t))=-e^{i(\omega_{\bk,1}+\omega_{\bk,2})t}\,\frac{k_z\left(m_1+\omega_{\bk,1}-m_2-\omega_{\bk,2}\right)}{2\sqrt{\omega_{\textbf{k},1}\omega_{\textbf{k},2}(m_1+\omega_{\textbf{k},1})(m_2+\omega_{\textbf{k},2})}}\,,\\[2mm]
V^{22}_\textbf{k}(t)&\equiv&(u^{2\dagger}_{\bk,1}(t),v^2_{-\bk,2}(t))=-V^{11}_\textbf{k}(t)\,,\\[2mm]
V^{12}_\textbf{k}(t)&\equiv&(u^{1\dagger}_{\bk,1}(t),v^2_{-\bk,2}(t))=-e^{i(\omega_{\bk,1}+\omega_{\bk,2})t}\,\frac{(k_x-i k_y)\left(m_1+\omega_{\bk,1}-m_2-\omega_{\bk,2}\right)}{2\sqrt{\omega_{\textbf{k}1}\omega_{\textbf{k},2}(m_1+\omega_{\textbf{k},1})(m_2+\omega_{\textbf{k},2})}}\,,\\[2mm]
V^{21}_\textbf{k}(t)&\equiv&(u^{2\dagger}_{\bk,1}(t),v^1_{-\bk,2}(t))=-e^{i(\omega_{\bk,1}+\omega_{\bk,2})t}\,\frac{(k_x+i k_y)\left(m_1+\omega_{\bk,1}-m_2-\omega_{\bk,2}\right)}{2\sqrt{\omega_{\textbf{k},1}\omega_{\textbf{k},2}(m_1+\omega_{\textbf{k},1})(m_2+\omega_{\textbf{k},2})}}\,,\\[2mm]
\nonumber
U_\textbf{k}(t)&\equiv&(v^{s\dagger}_{-\bk,2}(t),v^r_{-\bk,1}(t))=(v^{s\dagger}_{\bk,2}(t),v^r_{\bk,1}(t))\\[2mm]
&=&e^{i(\omega_{\bk,1}-\omega_{\bk,2})t}\,\frac{\left[|\textbf{k}|^2+\left(m_1+\omega_{\bk,1}\right)\left(m_2+\omega_{\bk,2}\right)\right]}{2\sqrt{\omega_{\textbf{k},1}\omega_{\textbf{k},2}(m_1+\omega_{\textbf{k},1})(m_2+\omega_{\textbf{k},2})}}\,\delta_{rs}\,.
\label{uk}
\end{eqnarray}
We notice that the above coefficients are non-vanishing, 
since (anti-)particle spinors with different masses
are not orthogonal to each other. 
From the above relation, we also derive
\be
V^{r1}_{\textbf{k}}(t)\,V^{s1\,*}_{\textbf{k}}(t)+V^{r2}_{\textbf{k}}(t)\,V^{s2\,*}_{\textbf{k}}(t)=|V_{\textbf{k}}|^2\delta_{rs}\,,
\ee
where
\be
|V_{\textbf{k}}|^2=\left[\left(\frac{\omega_{\bk,1}+m_{1}}{2\omega_{\bk,1}}\right)^{\frac{1}{2}}
\left(\frac{\omega_{\bk,2}+m_{2}}{2\omega_{\bk,2}}\right)^{\frac{1}{2}}\left(\frac{|\bk|}{\omega_{\bk,2}+m_{2}}-\frac{|\bk|}{\omega_{\bk,1}+m_{1}}\right)\right]^2\,.
\label{eqn:Vk2}
\ee
It is straightforward to show that
\be
\label{upv}
|U_\bk|^2+|V_\bk|^2=1\,,
\ee
which guarantees that the flavor operators~\eqref{eqn:annihilator}
and~\eqref{annihilatorb} and their hermitian conjugates are
still canonical (at equal times)~\cite{BV95}. 
 
The appearance of a nontrivial
Bogoliubov transformation in Eqs.~\eqref{eqn:annihilator}
and~\eqref{annihilatorb} makes it clear that the flavor
annihilators $a^r_{\bk,e}$ and $b^r_{\bk,e}$ do not actually
annihilate the mass vacuum. In other terms, the vacua
for flavor and mass fields (and the related Fock spaces)
are not equivalent to each other, the former becoming 
a condensate of entangled massive particle/antiparticle pairs with density
\be
\label{dens}
_{e,\mu}\langle0(\theta,t)|a^{r\dagger}_{\bk,i}a^r_{\bk,i}|0(\theta,t)\rangle_{e,\mu}=\sin^2\theta |V_\bk|^2\,,\,\forall t\,,\qquad i=1,2\,.
\ee
This is even more evident if we look at the
explicit expression of $|0(\theta,t)\rangle_{e,\mu}$
in terms of $|0\rangle_{1,2}$. For instance, in the reference
frame where $\textbf{k}=(0,0,k)$, this reads
\begin{eqnarray}
\label{vacstr}
\non
|0(\theta,t)\rangle_{e,\mu}&\hspace{-1mm}=\hspace{-1mm}& \prod_{{k}}\prod_{r}
\Big[\left(1-\sin^{2}\theta\;|V_\bk|^{2}\right)
-\varepsilon^{r}\sin\theta\,\cos\theta\, V_\bk
\,(a^{r\dagger}_{{\bk},1}\,b^{r\dagger}_{-{\bk},2}
+a^{r\dagger}_{{\bk},2}\,\beta^{r\dagger}_{-{\bk},1})  
\\[2mm] 
&&+\,\varepsilon^{r}\sin^{2}\theta\,V_\bk\,(U_\bk^*\,\alpha^{r\dagger}_{{\bk},1}\,b^{r\dagger}_{-{\bk},1}-U_\bk\,
a^{r\dagger}_{{\bk},2}\,b^{r\dagger}_{-{\bk},2})
+\sin^{2}\theta\,V_\bk^{2}\,
a^{r\dagger}_{{\bk},1}\,b^{r\dagger}_{-{\bk},2} \,
a^{r\dagger}_{{\bk},2}\,b^{r\dagger}_{-{\bk},1}
\Big]|0\rangle_{1,2}\,,
\end{eqnarray}
where $\varepsilon^{r}=(-1)^r$ and
we have omitted for simplicity the time-dependence 
in the r.h.s. This relation shows that the flavor vacuum
is made up of pairwise correlated particles/antiparticles of both equal and different masses and opposite momenta. 
It is worth noting that the presence of different types of pairs
makes this state different from the ground state of the BCS theory of superconductivity~\cite{Bardeen:1957kj}, where only one type of pair is involved. 

Again, the density~\eqref{dens} vanishes for $\theta=0$ and/or $m_2=m_1$, thus giving back the expected equality between the flavor and mass representations when there is no mixing. The same occurs 
in the limit of large momenta with respect
to the masses $m_1$ and $m_2$, consistently with 
the recovery of the standard quantum mechanical picture. 

Of course, by exploiting the symmetric structure of Eq.~\eqref{aflav}, 
one can reverse the above reasoning and study
the inherent structure of the mass vacuum in terms of the definite flavor quanta. What is obtained in that case is that
such state would appear as a condensate of particle/antiparticle pairs of both equal and different flavors. For more details
on the QFT formalism of field mixing and the related physical implications, we refer the reader to~\cite{BV95,3f,Cabo,Vacent}.
Specifically, in~\cite{Cabo} it has been shown that the flavor Fock space cannot be obtained by the direct product of the Fock spaces
for massive fields. Therefore, the nontrivial nature of QFT mixing is a feature boiling down to the
nonfactorizability of the flavor states in terms of those with definite massses, including the vacuum. The entanglement content
of the flavor vacuum has been quantified in~\cite{Vacent} 
for the case of boson mixing and
in the limit of relatively small difference of masses. 
 
\section{Modified Unruh distribution for mixed fields}
\label{Int} 
In this Appendix we give more details on the calculation
of the integral~\eqref{nfg}
\be
\mathcal{N}^{FG}_{j,\kappa,\sigma}\equiv\sum_{r=1,2}\int dk_x \,F_{j,r,\sigma}(k_x,\Omega)\left[G_{j,1,\sigma}^{\hspace{0.1mm}*}(-k_x,\Omega)\,V^{r1}_\bk\,+\,G_{j,2,\sigma}^{\hspace{0.1mm}*}(-k_x,\Omega)\,V^{r2}_\bk
\right]U_\bk\,.
\ee
for spin $j=1$ and flavor $\ell=e$ (clearly, similarly considerations
hold for $j=2$ and/or $\ell=\mu$). By setting $\epsilon\equiv\sqrt{m_2^2-m_1^2}/\omega_{\bk,1}$ and expanding 
around $\epsilon\ll1$, to the leading order one can prove that
\be
\label{b2}
\mathcal{N}^{FG}_{1,\kappa,1}\simeq\int dk_x\,\epsilon^2\, \mathcal{I}_\kappa(k_x)\,,
\ee
where
\begin{eqnarray}
\non
\mathcal{I}_\kappa(k_x)&=&\frac{i\,e^{-\pi\Omega}}{64\,m_1\mu_{\vec{k},1}\omega_{\bk,1}^{\hspace{0.2mm}4}}
\left(\frac{\omega_{\bk,1}+k_x}{\omega_{\bk,1}-k_x}\right)^{i\hspace{0.2mm}\Omega}
\left(1+\frac{|\bk|^2}{(m_1+\omega_{\bk,1})^2}\right)\\[2mm]
\non
&&\times\Bigg\{k_z^2\left(\mu_{\vec{k},1}+(\omega_{\bk,1}+k_x)\sqrt{\frac{\omega_{\bk,1}+k_x}{\omega_{\bk,1}-k_x}}\right)
\bigg[i(k_x-ik_y)\left(\mu_{\vec{k},1}+(\omega_{\bk,1}+2m_1-k_x)\sqrt{\frac{\omega_{\bk,1}-k_x}{\omega_{\bk,1}+k_x}} \right)\\[2mm]
\non
&&+\Big[\mu_{\vec{k},1}(k_x-ik_y+m_1+\omega_{\bk,1})+
\sqrt{\frac{\omega_{\bk,1}-k_x}{\omega_{\bk,1}+k_x}}\Big(k_z^2+(k_y-im_1)\big(ik_x+k_y-i(m_1+\omega_{\bk,1})\big)\Big)\Big]\bigg]\\[2mm]
\non
&&+\Big[\mu_{\vec{k},1}(k_x-ik_y-m_1-\omega_{\bk,1})-
\sqrt{\frac{\omega_{\bk,1}+k_x}{\omega_{\bk,1}-k_x}}
\Big(k_y^2+k_z^2+k_x(ik_y+m_1)+ik_y\omega_{\bk,1}+m_1(m_1+\omega_{\bk,1})\Big)\Big]\\[2mm]
\non
&&\times \Big[-ik_z^2\left(\mu_{\vec{k},1}+(\omega_{\bk,1}+2m_1-k_x)\sqrt{\frac{\omega_{\bk,1}-k_x}{\omega_{\bk,1}+k_x}}
\right)+(k_x+ik_y)\bigg(\mu_{\vec{k},1}(k_x-ik_y+m_1+\omega_{\bk,1})\\[2mm]
&&+\sqrt{\frac{\omega_{\bk,1}-k_x}{\omega_{\bk,1}+k_x}}\Big(k_z^2+(k_y-im_1)\big(ik_x+k_y-i(m_1+\omega_{\bk,1})\big)\Big)
\bigg)\Big]
\Bigg\}\,.
\label{b3}
\end{eqnarray}
Due to the nontrivial dependence of $\mathcal{I}_\kappa$
by $k_x$, the integral~\eqref{b2} cannot be solved
analytically. However, for our purposes it is enough to note
that $\mathcal{N}^{FG}_{1,\kappa,1}$ can be rewritten
equivalently as
\be
\label{b4}
\mathcal{N}^{FG}_{1,\kappa,1}\simeq \frac{|\Delta m^2|}{\mu_{\vec{k},1}^2}\,e^{-\pi\Omega}\,\mathcal{H}(\mu_{\vec{k},1})\,, 
\ee
where we have separated out the dependence
on both the mass difference $|\Delta m^2|$ and the Rindler energy $\Omega$. 
$\mathcal{H}(\mu_{\vec{k},1})$ is a shorthand notation for the (dimensionless) $k_x$-integral 
of the residual function\footnote{Strictly speaking, 
$\mathcal{H}(\mu_{\vec{k},1})$ also depends on $\Omega$. Nevertheless, 
since the dependence is through an oscillating term, it does not affect significantly the outcome of the $k_x$-integral and can be in principle neglected.}. 

The integral~\eqref{b2} can be estimated numerically. In the limit of relativistic
neutrinos and for values of $\Omega$ such that the unmodified  
Fermi-Dirac spectrum in Eq.~\eqref{moddi} differs appreciably from zero, 
$\mathcal{N}^{FG}_{j,\kappa,\sigma}$ is finite and its real part assumes
positive values. As $\Omega$ increases, the rapidly oscillating behavior of 
$\mathcal{H}(\mu_{\vec{k},1})$ makes it difficult even numerical evaluations.
However, this regime turns out to be less relevant, 
since the larger the energy $\Omega$, the lower the expected number
of particles in the vacuum condensate. Further investigation
of these technicalities is under active consideration
and will be carried out elsewhere.

\end{document}